\documentclass[a4paper]{article}

\usepackage[english]{babel}
\usepackage[utf8]{inputenc}
\usepackage{amsmath}
\usepackage{amsfonts}
\usepackage{amssymb}
\usepackage{algorithm}
\usepackage{algorithmic}
\usepackage{multirow}
\usepackage{graphicx}
\usepackage{caption} 
\usepackage{rotating}
\providecommand{\keywords}[1]
{
  \small	
  \textbf{\textit{Keywords---}} #1
}

\title{Information Consumption and Boundary Spanning in Decentralized Online Social Networks: the case of Mastodon Users\thanks{This is a preprint version of the article \cite{osnem} published with Online Social Networks and Media, vol. 30:100220, June 2022. Elsevier.}}

\author{Lucio {La Cava}   \and Andrea Tagarelli}

\date{\small Dept. Computer Engineering, Modeling, Electronics, and Systems Engineering (DIMES), University of Calabria, 87036 Rende (CS), Italy\\[1ex] \{lucio.lacava,   tagarelli\}@dimes.unical.it }

\begin{document}
\maketitle

\begin{abstract}

Decentralized Online Social Networks (DOSNs) represent a growing trend in the social media landscape, as opposed to the well-known centralized peers, which are often in the spotlight due to privacy concerns and a vision typically  focused on monetization through user relationships. By exploiting open-source software, DOSNs allow users to create their own servers, or instances, thus favoring the proliferation of platforms that are independent yet interconnected with each other in a transparent way. 
 Nonetheless, the resulting cooperation model, commonly known as the Fediverse, still represents a world to be fully discovered, since existing studies  have mainly focused on a limited number of  structural aspects of interest in DOSNs. In this work, we aim to fill  a lack of study on user relations and roles in DOSNs, by taking two main actions:   understanding the impact of decentralization on how users relate to each other   within their membership  instance and/or across different instances, and unveiling  user roles that can   explain two interrelated axes of social behavioral phenomena, namely information consumption and boundary spanning.   
To this purpose, we build our analysis on user networks from Mastodon,  since it represents the most widely used DOSN platform.  

We believe that the findings drawn from our study on Mastodon users' roles and information flow can pave a  way for further development of  fascinating research on DOSNs.

\end{abstract}
\hspace{10pt}

\vspace{5mm}
\section{Introduction}
\label{sec:introduction}
The  realm of online social networks (OSNs) in the Internet landscape has  determined  a radical metamorphosis in our lives.  
Over the past two decades, we observed the advent and the rapid growth of numerous OSN platforms. Their extensive diffusion across the globe and in our lives dramatically changed how we share  information  and socialize with each other. These platforms introduced new paradigms and imposed new constraints on the way we communicate within their boundaries.

However, this   has eventually exhibited potentially worrying aspects. Since centralized OSN platforms --- i.e., hosted and controlled by unique owners --- have become pervasive, they started leveraging social media marketing strategies,  which exploit  targeted advertisements and content personalization to maximize their purposes. Although these choices would presumably increase user engagement and favor the businesses of these platforms, latent side effects are around the corner. Just to mention, there is a risk of running into ``information bubbles'' or ``echo chambers'', with related   distortions of social reality. Furthermore, privacy and security concerns might also arise when a single actor handles all user data. Hence, the need for alternatives is evident: social platforms were born to interconnect people, whereas nowadays, they mainly end up doing marketing by exploiting these connections. 

\textit{Decentralized Online Social Networks} (DOSNs) have emerged   in response to the aforementioned demanding issues, so as to bring the user back to the center of the social stage and to support spontaneous interactions, no longer biased by marketing mechanisms. Decentralization allows going beyond a single owner, hence giving greater privacy control to the users. The main components to pursue these objectives are the availability of open-source software that users can quickly adopt, and the possibility of connection among different servers --- commonly  referred to as \textit{instances} --- on which this software runs, which is achieved by leveraging specific communication protocols. 
This seamless connection between instances leads to a \textit{federated} model, which allows a user registered to an instance to interact with other users on other instances in a completely natural and transparent way --- i.e., without the need for making further subscriptions --- similarly to email services. The development of countless platforms based on this federated model has led to the emergence of the so-called federated universe, commonly known as the \textit{Fediverse}, a widespread social network made up of many instances inherently interconnected with each other.  
These include  \textit{Mastodon} and \textit{Pleroma} for microblogging, \textit{Pixelfed} and \textit{Peertube} for image and video hosting, resp.,   \textit{Funkwhale} for audio hosting, and others. 

 Mastodon is by far the  most popular platform in the Fediverse and the one that has attracted the most attention by the research community.  
   Mastodon is a microblogging platform that offers a user experience in line with Twitter, under many aspects, while including original and valuable functionalities. For instance, Mastodon users can publish content (dubbed \textit{toots}) and share other people's content via the \textit{boost} function (similar to the \textit{retweet} in its centralized counterpart, Twitter). Furthermore, along with these ``traditional'' services, the decentralized nature of Mastodon favors the formation of interest-based communities (i.e., individual instances) analogously to what happens in other OSNs such as Reddit. This feature results in an enhanced content management on Mastodon. On the one hand, users can declare some content as inappropriate for a given instance using the \textit{content warning} feature, while offering a textual complement of such content (i.e., a \textit{spoiler}); on the other hand, instances' administrators can explicitly declare the topics of interest that characterize their instances, prohibit some types of contents or even close registrations for their instances. It should be noted that the latter feature only affects the subscription of new users at a specific instance and does not impact on the possibility of interacting with such instances, thanks to the interoperability ensured by the underlying protocols.
 
From a   technical point of view, 
Mastodon adopts the \textit{ActivityPub} protocol,\footnote{https://www.w3.org/TR/activitypub/} which provides client-to-server and server-to-server communication capabilities.   Combined with subscription mechanisms aimed at retrieving information from remote instances,  this protocol supports the aforementioned  seamless communication between users of different Fediverse instances. The outcomes of this extended followship mechanism on the user experience are manifold. The most noteworthy one is the subdivision of the user timeline into three possible levels of abstraction: \textit{home}, \textit{local}, and \textit{federated} timelines. Specifically,   the home timeline contains toots generated by the followed users, whereas the local and federated timelines contain toots created within the home instance and public toots from all (local or remote) users known to it,  respectively.

\paragraph{\bf Related work.\ } 
Although recently emerged, decentralized social platforms and their novelties have attracted a certain  attention from researchers of various disciplines.   Datta et al.~\cite{Datta2010} investigated the motivations concerning the decentralization of online social networking, exploring various raised challenges and opportunities. Guidi et al.~\cite{Guidi2018} analyzed DOSNs focusing on data management and availability, information diffusion, and privacy; also, they examined limitations and problems associated with the decentralized paradigm.

As previously mentioned, Mastodon  has gained particular consideration from the research community~\cite{Cerisara2018,Trienes2018,Zignani2018,Raman2019,Zignani2019,Zulli2020}. The qualitative interview-based analysis conducted by Zulli et al.~\cite{Zulli2020} sheds light on how Mastodon enables content diversification and community autonomy, supporting horizontal growth between instances rather than vertical growth within instances. 
Zignani et al.~\cite{Zignani2018,Zignani2019} leveraged network analysis to explore the interactions between Mastodon users through a set of structural statistics such as degree distribution, triadic closure, and assortativity. They used these results to compare Mastodon with the platform closest to it from a user experience perspective, i.e.,  Twitter~\cite{Zignani2018}. According to the in-degree and out-degree distributions  analyzed by Zignani et al., Mastodon exhibits a more balanced distribution between followers and followees (with differences ranging between -250 and 250) than Twitter, and  a limited presence of bots (around 5\%) w.r.t. Twitter (where their fraction of user base was found to be around 15\%~\cite{Varol2017}).  
 Moreover, when focusing on the network formed by reciprocated edges only, the authors found that the local clustering coefficient of nodes stands numerically between the one of Twitter and the one of Facebook. Mastodon also diverges from well-known centralized OSNs in terms of assortativity, as indicated by the lack of correlation between source and destination out-degree, and between source out-degree and destination in-degree. In addition, negative correlation (-0.1) was observed between source and destination in-degrees, which indicates that a user's popularity is inversely proportional to the users s/he follows.    Zignani et al. also evaluated the influence of decentralization on relationships between users~\cite{Zignani2019}, revealing how each instance has its own footprint that impacts on the way users connect with each other.

In~\cite{LaCava2021}, we recently  contributed  to studying  the decentralized paradigm through Mastodon from a different perspective, i.e., the \textit{instance} level. 
First, we created an updated and highly representative dataset of the relationships between Mastodon users, upon which we inferred a network model  capturing the   relationships between Mastodon instances.  Leveraging this model, we delved into the main structural characteristics of Mastodon, taking a macroscopic as well as a mesoscopic perspective, also analyzing the Mastodon instance network backbone.  Our earlier study revealed a fingerprint characterizing the network of Mastodon instances, through which we distinguish Mastodon from the most widely used centralized OSNs. Also, we spotted the development of a mutual-reinforcement mechanism between instances to reduce the potential sectorization bias deriving from the decentralization, which is a feature further strengthened by an emerging modular structure between the instances. In this regard, we provided insights into the nature of the communities of Mastodon instances, unveiling the main factors that influence their development, e.g., topics, languages, and temporal processes. We also shed light on the linkage mechanism  between   instances which shows  a negative degree correlation, thus revealing   how users tend to interact regardless of the relevance of their ``home'' instance. Finally, we evaluated the evolution of the platform and its role within the Fediverse during the last few years, assessing the achievement of its structural stability and the temporal consolidation of the role of the most relevant instances.

It is worth noticing that, being developed on an up-to-date crawling of Mastodon corresponding to a much larger network  than in previous studies, our earlier work in~\cite{LaCava2021}  has set the current state-of-the-art data for modeling Mastodon; however, despite the in-depth investigation made on this data, all findings drawn from our earlier  study are at instance level only.

\paragraph{\bf Contributions.\ } 
Our research work hence aims to fill  a lack of study on user relations and roles in DOSNs, and in this respect we want to pursue two main interrelated goals: 
\begin{itemize}
    \item First, since DOSNs embrace a myriad of  instances, we are interested in understanding whether and to what extent interesting, decentralization-driven user behaviors arise within the membership instances, across different instances, or even correspond to  mixed behaviors. This might favor the comprehension and the modeling of the information flow within and between multiple instances. 
    \item Second, since the human-centric approach of DOSNs undervalues   artificially imposed interactions, such as those deriving from boosting or advertisement mechanisms, we want to assess how users shape their roles in a more spontaneous social networking context. 
    In this respect, our focus is on   user roles that are essential to explain two interrelated axes of behavioral phenomena in online social networks, namely information consumption and boundary spanning. 
    The dualism between information consumption and boundary spanning can indeed profoundly affect the scope of the information flow within a network and its fluidity, e.g., whether information flows rapidly and spreads widely or remains confined to specific areas of the network.
\end{itemize}

 To the best of our knowledge, no works have been proposed so far  to analyze user roles and behaviors in DOSNs based on the above aspects.
 Note also that our perspectives on the aforementioned aspects of interest in this work  are totally independent  from knowledge about textual or media contents produced and exchanged  through the Fediverse, thus exploiting  only the topological information of the user relation network.

 To conduct our research study, we shall  focus on the most widely known  and representative Fediverse platform, i.e., Mastodon, which is hence the best suited  as case in point for investigating on the DOSN landscape. Moreover, this also allows us to capitalize on  up-to-date data resources and relating findings from our earlier work~\cite{LaCava2021}.

Our roadmap to delve into the understanding of the above discussed aspects  will be developed so as to pursue a number of objectives that can be summarized into the following research questions:
 
\begin{enumerate}
\item[\textbf{Q1} --] \textit{The User Network structure}: What are the main structural characteristics of the network of following relations between Mastodon users?

\item[\textbf{Q2} --] \textit{Representative instances}: Are the users belonging to the most relevant Mastodon instances representative of the entire user network?

\item[\textbf{Q3} --] \textit{Boundaries and bridges}: Are Mastodon users involved in inter-instance links and how do they act as local bridges?  

\item[\textbf{Q4} --] \textit{Over-consumption}: 
Are there Mastodon users who tend to over-consume, i.e., lurk, others' information? 
Is this behavior bounded to the membership instances    or it spans across the instance boundaries?

\item[\textbf{Q5} --] \textit{Dual role users}: Are there users who behave as both lurkers and bridges within their own instance?

\item[\textbf{Q6} --] \textit{Alternate role users}: Can user behavior vary according to the observation scale? That is, can a user be a lurker within her/his instance and simultaneously act as a bridge between instances, or vice versa?
\end{enumerate}

\paragraph{\bf Plan of the paper.\ } 
The remainder of the paper is organized as follows. 
Section~\ref{sec:data} introduces to the Mastodon data and the network models we used in our analysis.   
Section~\ref{sec:network-analysis} presents our structural analysis of the Mastodon user networks, by first considering  the full set of users and their relations, then focusing on  a representative subset of the user network corresponding to a selection of the most relevant instances in Mastodon.  
Section~\ref{sec:behavioral-analysis} provides insights into user roles that are relevant to  boundary spanning and information consumption behaviors for the users in Mastodon. 
 Section~\ref{sec:discussion}   summarizes the main lessons learned from our analysis,  while Section~\ref{sec:conclusions} concludes the paper and provides pointers for future research.

\section{Data Extraction and Network Modeling}
\label{sec:data}
In our earlier work~\cite{LaCava2021}, we   developed a privacy-friendly crawler upon the publicly available Mastodon REST APIs~\footnote{https://docs.joinmastodon.org/api/} to build an up-to-date and highly representative dataset.  
It is worth emphasizing that, to preserve   privacy   requirements,   we relied on authenticated requests only, i.e., those towards the instances that allowed accountable requests through their APIs, and   we also avoided using any scraping tools.

To account for the decentralized nature of Mastodon, we leveraged on the \textsl{instances.social} website,\footnote{https://instances.social/} which politely keeps track of the Mastodon instances panorama. In particular, this website enabled us to locate some ``seed'' instances (i.e., the online ones at the time of crawling) from which we started our exploration of the Mastodon Fediverse. By getting information on the timelines of about 900 instances, we reached more than 80\,000 users, who represented the starting point of a \textit{breadth-first-search} to discover new connections and, consequently, more     users.  
In this regard, we point out that although the toots (delivered over the timelines of the   seed instances) were inspected to discover the corresponding users, the toot data was never stored, 
  therefore \textit{our study described in this work is totally agnostic of textual contents}. 
  Furthermore,  the interactions between users in terms of incoming and outgoing links were anonymized through proper hashing functions at the time of their acquisition. 

After processing the fetched data, we came up with about 1.4M unique users and 18M unique links between them, traversing more than 16\,000 instances. The protocol underlying  Mastodon, i.e., ActivityPub, supports seamless communication between all the Fediverse platforms. This implies that  the data obtained by means of the APIs can in principle also concern interactions between Mastodon instances and instances pertaining to other services in the Fediverse. Therefore, using the aforementioned \textsl{instances.social} and the \textsl{fediverse.party} platforms,\footnote{https://fediverse.party/en/mastodon} we discerned the Mastodon instances within our dataset, splitting them into online and temporarily offline ones, where the latter correspond to instances that   keep an inactive status for at most two weeks (according to the instances documentation provided by   \textsl{instances.social}).
 As a result, we discovered 6\,960 Mastodon instances, among which 1\,116 were online.  
 
Upon the extracted data, we built networks whose entities (i.e., nodes) represent users, and we modeled their relationships either at the level of the whole Mastodon network or at the level of individual instances.    
Let us denote with $\mathcal{U}$  the set  of users and with   $\mathcal{I}$ the set of instances available in the   extracted Mastodon data. 
We define the \textit{Mastodon user network} as a   directed graph    $\mathcal{G} = \langle \mathcal{V}, \mathcal{E} \rangle $,  where the node set $\mathcal{V}$ contains pairs $(u,i)$,  with $u \in \mathcal{U}$ and $i \in \mathcal{I}$,  and the edge set $\mathcal{E} \subseteq \mathcal{V} \times \mathcal{V}$ corresponds to the set of \textit{following} relations,   such that any $(x, y) \in \mathcal{E}$ with $x=(u,i)$ and $y=(v,j)$  means that user $u$ in instance $i$ follows user $v$ in instance $j$;  note that $u$ may coincide with $v$ only if  $i\neq j$. 

Given a target instance $i \in \mathcal{I}$, we define the \textit{instance-specific user network}  of  $i$ as the directed subgraph $G_i=\langle V_i, E_i\rangle$ induced from $\mathcal{G}$, such that $V_i= \{u \ | \ (u,i) \in \mathcal{V}\} \subseteq \mathcal{U}$ and $E_i$ is the set of edges $(u,v)$ with $u$ following $v$ in instance $i$.

Given a set of target instances $\mathcal{M} \subset \mathcal{I}$, we define the network   of the relations between users of the instances $\mathcal{M}$, dubbed \textit{merged network},  as the directed subgraph $G_{\mathcal{M}} =\langle V_{\mathcal{M}}, E_{\mathcal{M}}\rangle$ induced from $\mathcal{G}$, such that $V_{\mathcal{M}}= \{u \ | \ (u,i) \in \mathcal{V}\land i \in \mathcal{M}\} \subseteq \mathcal{U}$ and $E_{\mathcal{M}}$ is the set of edges $(u,v)$ with $u$ following $v$   in some instance  of $\mathcal{M}$.

\section{User Network Structure}
\label{sec:network-analysis}

In this section we begin with answering our first research question (\textbf{Q1}), i.e., understanding the main structural traits of the user network in Mastodon (Section~\ref{sec:full-network}).  Next, we take the opportunity of investigating on the presence of noisy or irrelevant user relations according to  requirements  specified for their membership instances (Section~\ref{sec:filtered-network}). 
 This   eventually leads us to answer our second research question (\textbf{Q2}) by identifying and analyzing the subset of the user network corresponding to the most relevant instances in Mastodon (Section~\ref{sec:merged-network}), which will be used as our workbench for the subsequent user behavioral analysis.

\begin{table}[t!]
\centering
\caption{Main structural characteristics of the user networks derived from Mastodon.}
\label{tab:networks-stats}
\rmfamily
\scalebox{0.85}{
\begin{tabular}{|l||c||c||c|}
\hline
& \textit{User network} & \textit{User network} &  \textit{Top-5 instance}\\
& \textit{\small{(full)}}  & \textit{\small{(filtered)}}   &  \textit{merged}\\
&  &  & \textit{user network} \\
\hline\hline
\#nodes & 1\,315\,739 & 1\,237\,985 & 657\,712 \\ 
\#edges & 17\,252\,347 & 16\,239\,329 & 8\,227\,553 \\
density   &   1e-05 & 1e-05 & 2e-05  \\
\hline\hline
\% sources  &   34.2\% & 34.7\% & 29.7\% \\
\% sinks   & 7.5\% & 6.7\% & 3.1\% \\
average degree$^*$ &  21.925 & 22.007 & 21.445 \\
average in-degree   &  13.112 & 13.118 & 12.509 \\
{degree assortativity}$^*$ & -0.040 & -0.042 & -0.072 \\
{degree assortativity} & -0.032 & -0.033 & -0.048 \\
\hline \hline
transitivity$^*$   &   0.003 & 0.003 & 0.002  \\
clustering coefficient$^*$   & 0.393 & 0.398 & 0.401  \\
clustering coefficient {\scriptsize \textit{(full averaging)}}$^*$   &  0.315 & 0.323 & 0.357  \\
reciprocity   &  32.8\% & 32.2\% & 28.6\%  \\
\hline\hline
average path length   & 5.326$^{\dag}$ & 5.312$^{\dag}$ & 5.088 (5.162$^{\dag}$) \\

\#strongly connected components    &   565\,865 & 526\,349 & 223\,935  \\
\#weakly connected components$^*$   &   327 & 320 & 141 \\
\hline \hline 
modularity {\scriptsize by \textit{Louvain}} & 0.737 & 0.736 & 0.688 \\ 
\#communities {\scriptsize by \textit{Louvain}} & 580 (125) & 578 (111) & 384 (89) \\ 

modularity {\scriptsize by \textit{Louvain}}$^*$ &  0.717 & 0.717 & 0.658 \\ 
\#communities {\scriptsize by \textit{Louvain}}$^*$ &  565 (127) & 518 (109) & 412 (90)\\ 

modularity {\scriptsize by \textit{Leiden}}$^*$ & 0.743 & 0.741 & 0.688 \\ 
\#communities {\scriptsize by \textit{Leiden}}$^*$ & 629 (138) & 629 (127) & 417 (97) \\ 

\#communities {\scriptsize by \textit{Infomap}} & 594 (53) & 568 (52) & 273 (48) \\ 
\#communities {\scriptsize by \textit{Infomap}}$^*$ & 359 (19) & 349 (19) & 178 (17) \\ 

\hline
\end{tabular} 
}
\vspace{1mm}
{\footnotesize
\begin{flushleft} 
$^*$ Statistic calculated by discarding the edge orientation \\
$^{\dag}$ Statistic calculated as $ln(N)/(ln ln(N))$, where $N$ denotes the number of nodes  \\ 
\end{flushleft}
}
\end{table}

\subsection{The Mastodon user network}
\label{sec:full-network}
As our first action towards the understanding of  behaviors of Mastodon users, we      gained a comprehensive view of the main features of the Mastodon user network, both from a macroscopic and mesoscopic perspective. 

In the first data column of   
Table~\ref{tab:networks-stats}, we report the main structural characteristics analyzed.  First, it stands out the  high numbers of nodes and edges available in our network,  which indeed captures  a large fraction  of the existing Mastodon user base.\footnote{At the time of writing of this work, we achieve a coverage that ranges between 62\% and 78\% of the full audience of Mastodon, according to  the \textsl{fediverse.party} and \textsl{instances.social} websites, respectively.} 
 We also  notice a reasonable fraction of source nodes (i.e., users having only outgoing links), which would include newcomers at the time of the data acquisition, and in general   users that take a peripheral role in the platform; moreover, 
 the fraction of sink nodes (i.e., users having only incoming links) is quite small, 
 which would indicate a moderate presence of users who do not appear to be interested in establishing or reciprocating connections with other Mastodon users.

Another helpful indicator concerning the linkage between  users is expressed by the degree correlation or \textit{degree assortativity}, i.e., the probability that a link between two nodes depends on their respective degrees~\cite{Newman2002,Newman2003}. 
Given the observed value, which is slightly negative yet close to zero, it happens that the Mastodon user network is an uncorrelated  network: this turns out to be explained due to a very interesting trait of Mastodon, since  the lack of degree correlation highlights how users relationships within Mastodon are generally driven by genuine interests, and     not biased by recommendation mechanisms within or across   instances.

We further delved into the relationships between users by investigating aspects of transitivity (i.e., the likelihood that two incident edges are completed by a third one, thus forming a triangle) and local clustering coefficient (i.e., how strongly connected are the neighbors of a node);   the latter was measured by  averaging either over all nodes (indicated as ``full averaging'' in Table~\ref{tab:networks-stats}) or only  nodes with degree greater than one. 
 The relatively higher local clustering coefficient w.r.t. the transitivity is not surprising,   given the  low density of the network. Moreover, when coupled with the moderate fraction of reciprocal edges (i.e., closed loops of length 2), this hints at the presence of strong local connectivity.
At the same time, the average path length is quite low, about 5.

To understand the community structure of the Mastodon user network, 
we resorted to the widely used \textit{Louvain}~\cite{Blondel2008} and \textit{Infomap}~\cite{Rosvall2008}  methods, along with the more recent \textit{Leiden} method~\cite{Leiden2019}. 
Louvain exploits a hierarchical greedy approach based on two phases, modularity optimization and community aggregation, which are repeated until there are no more changes to be made on the communities and a maximum of modularity is achieved.
Infomap optimizes the Map equation, which leverages the information-theoretic duality between finding community structures in a network and minimizing the description length of the movements of a random walker in a network.
  The Leiden method is designed to  improve upon the Louvain method, by providing guarantees on the   connectivity of the discovered communities through an iterative algorithm that includes   local moving and community aggregation stages, with the addition of an intermediate stage of refinement of the community connectivity. 
We used both the undirected and directed implementations of the Louvain\footnote{https://github.com/nicolasdugue/DirectedLouvain} and   Infomap\footnote{https://www.mapequation.org/infomap/} algorithms, whereas for the Leiden algorithm, we used the only available undirected implementation.\footnote{https://github.com/vtraag/leidenalg}

Table~\ref{tab:networks-stats} reports the number of communities found by each of the above algorithms, along with their modularity, if optimized by the corresponding method.  
 We distinguish between overall and meaningful communities, the latter being regarded as those  having at least ten users (enclosed in round brackets), and we indicate with * in the table the cases when we discard edge orientation. As shown by the modularity values above 0.7 for both Louvain and Leiden algorithms, the Mastodon user network appears to be characterized by a modular structure.  The number of communities found ranges between 359 (by undirected Infomap) and 629 (by  undirected Leiden), which however is significantly reduced when focusing on the  meaningful communities,  from   
  127  (by  undirected  Louvain) and 138 (by undirected Leiden) down to 19, resp. 53, by undirected, resp. directed, Infomap.

\subsection{Filtering out noisy instances}
\label{sec:filtered-network}
In our structural analysis of the Mastodon user network, we also considered to measure the effects of simplification of the network in terms of pruning of user relations  involving potentially noisy or irrelevant instances.  
To this purpose, here we capitalize  on related findings from our previous  study~\cite{LaCava2021},  in which we assessed the relevance of the instances according to the number of links they receive. In this regard, we observed statistical significance (based on a Kolmogorov-Smirnov test) of a lognormal fitting of the in-degree distribution when removing the instances that are pointed by  less than 51 other instances. 

Performing this pruning step on our Mastodon user network implies  the removal of approximately 100k nodes and 1M edges; nonetheless, it should be noticed that the main characteristics of the Mastodon user network structure have remained substantially unchanged. 
In fact, as shown in the central column of Table~\ref{tab:networks-stats}, all statistics are in line  with those relating to the original network. Such consistency allows us to argue that the most relevant instances strongly determine the backbone of the whole Mastodon user network, also  ensuring its robustness w.r.t. the removal of potentially noisy elements.

\begin{table}[t!]
\centering
\caption{Size of the top-5  relevant Mastodon instances and their aggregation (\textit{merged network}).}
\label{tab:subgraphs}
\begin{tabular}{|l||c|c|c|}
\hline
& \#nodes & \#edges & density\\
\hline \hline
\textsl{mastodon.social} & 305\,968 & 3\,408\,327 & 4e-05 \\
\textsl{pawoo.net} & 306\,753 & 4\,329\,562 & 5e-05 \\
\textsl{mastodon.xyz} & 16\,076 & 35\,631 & 1e-04 \\
\textsl{mstdn.io} & 16\,853 & 112\,805 & 4e-04 \\
\textsl{octodon.social} & 7\,082 & 34\,493 & 7e-04 \\
\hline\hline
\textit{merged network} & 657\,712 & 8\,227\,553 & 2e-05 \\
\hline
\end{tabular}
\end{table}

\subsection{Narrowing the focus: the top-5 instances}
\label{sec:merged-network}
We further investigated the impact of instance selection on the main traits of the Mastodon user network by focusing on the most important instances. 
Again following the lead of~\cite{LaCava2021}, we selected the  top-5 instances by relevance, according to the maximization of a threefold criterion based on number of registered users, number of involved links, and data access permission policy; this resulted in the following selection of instances:   \textsl{mastodon.social}, \textsl{pawoo.net}, \textsl{mastodon.xyz}, \textsl{mstdn.io} and \textsl{octodon.social}.\footnote{Note that, according to the \textsl{instances.social} website, the top five ranked instances might partly be changed, since   the time of our crawling of the Mastodon network,  w.r.t. one or all of the criteria we considered for the instance selection.}
 Table~\ref{tab:subgraphs} provides a   summary of their size information.  
 We then created the user network upon these instances according to the  \textit{merged network model} (defined in Section~\ref{sec:data}).

\begin{figure}[t!]
\centering
\begin{tabular}{cc}
\includegraphics[width=0.45\textwidth]{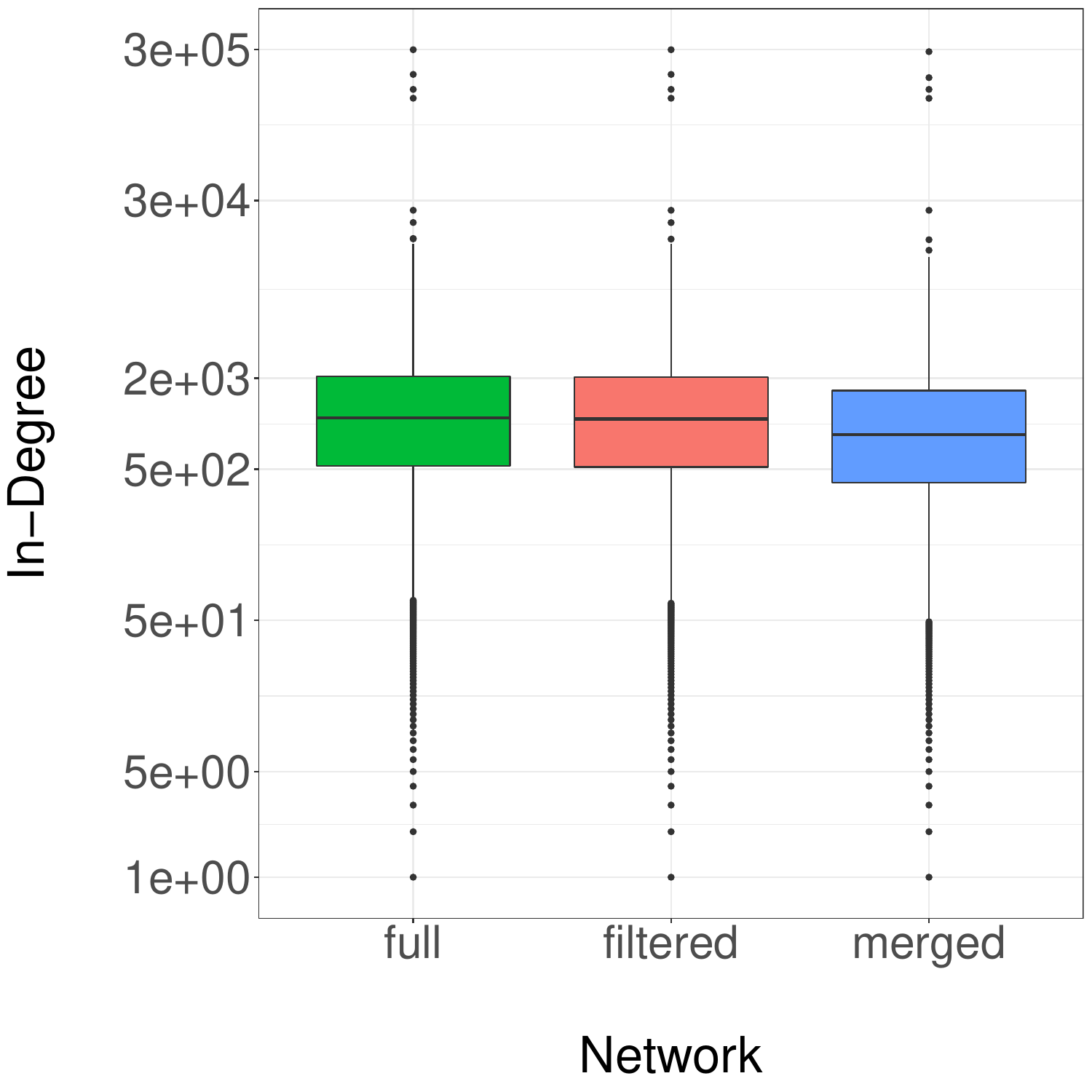} &
\includegraphics[width=0.45\textwidth]{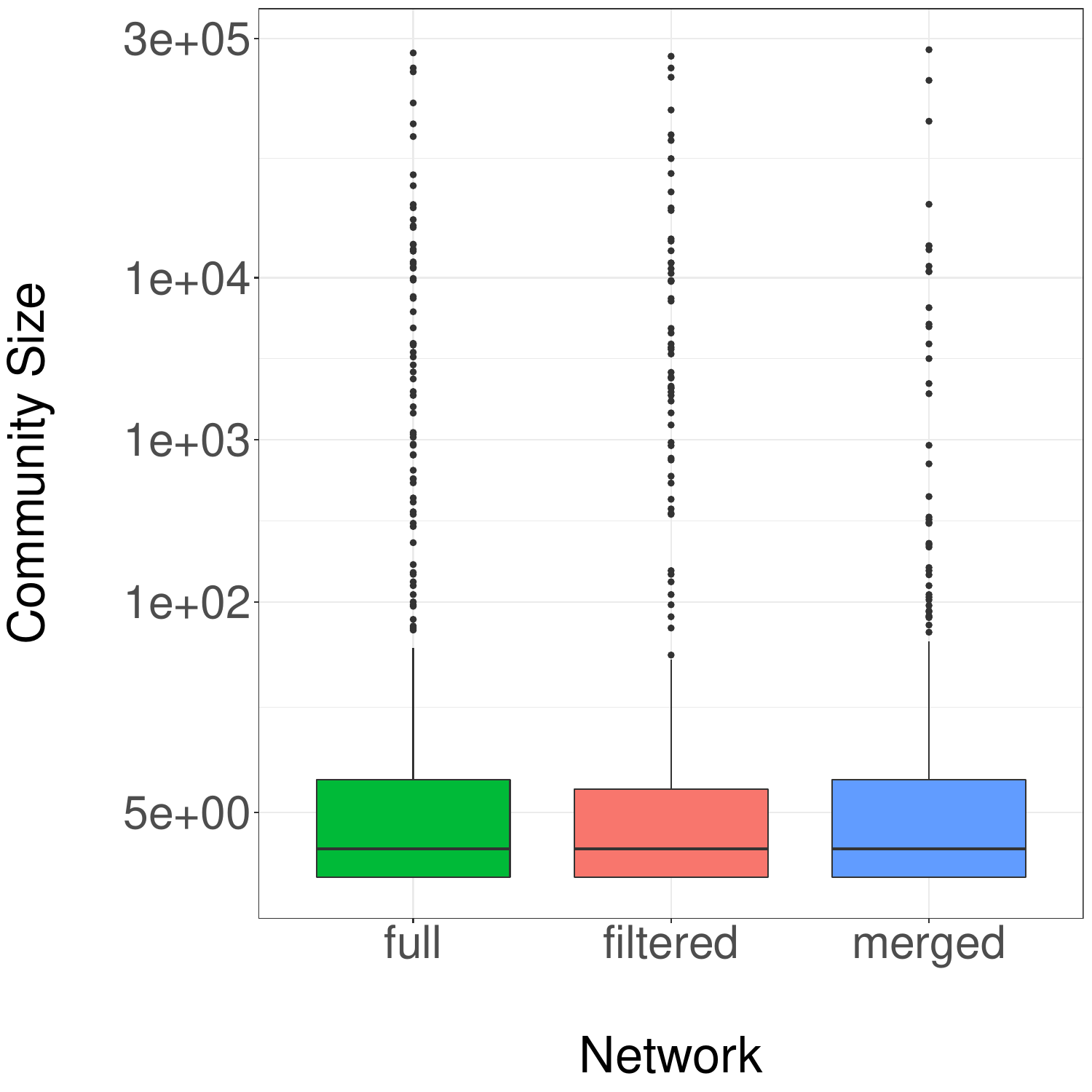}
\end{tabular}
\caption{In-degree distributions (on the left) and community size distributions  (on the right)   of the \textit{full}, \textit{filtered}, and top-5 instance \textit{merged} networks. (Community size distributions refer to the solutions by undirected Louvain)}
\label{fig:comparison}
\end{figure}

On this merged network, we replicated the structural analysis carried out on the whole and pruned  user-networks, in order to unveil whether and to what extent the user network of the top-5 instances  can be considered representative of the entire Mastodon user network. 
As shown in the rightmost  column of Table~\ref{tab:networks-stats}, the similarity between the statistics on the top-5 instance merged network and the corresponding ones of the  whole user network is evident, which      supports   our above  hypothesis of representativeness of the Mastodon user network. 
Note that, as concerns   the average path length,   we managed to calculate it exactly on the merged network, while for the much larger full and filtered networks  we were forced (due to computational  issues)   to an approximation generally valid for random scale-free uncorrelated networks  (cf. Table~\ref{tab:networks-stats}); in this respect, the approximated values of average path length appear to be very close to the exact value computed on the top-5 instance merged network.

To further strengthen our hypothesis, we also examined     the in-degree distributions and the community size distributions of the three user networks under evaluation.
As shown in Figure~\ref{fig:comparison}, for both in-degree and community size,  the three networks have     box plots that are very close to each other.  In particular, concerning the in-degree distributions,  the medians  are equal to 1\,092.5, 1\,078.5, and 846,  for the full, filtered, and merged networks, respectively; moreover, the median   of the community size distributions settles on 3 for all the considered networks. 
 This is remarkable, as not only confirms   the relatively low  relevance of the instances   removed from the entire user-network, but more interestingly, it indicates that the network of the top-5 instances can effectively be used as a  proxy for the   Mastodon user relations;  in addition, by focusing on a small number of large instances, this proxy can in principle enhance  our  interpretability of  the  behavioral patterns to discover in the Mastodon user relations.  

Upon the above  findings and remarks, we chose to narrow our focus on the top-5 instance merged network in the subsequent behavioral analysis of   Mastodon users.

\section{Boundaries, bridges, and over-consumption}
\label{sec:behavioral-analysis}

  In this section we delve into the relations between users in top-5 instance merged network in order to discover  user roles that are relevant in terms of two particular behavioral phenomena, namely  boundary spanning and information consumption. We elaborate on the former in Section~\ref{sec:boundaries} and on the latter in Section~\ref{sec:overconsumption}, which will lead us to answer our \textbf{Q3} and \textbf{Q4} research questions, respectively.

\subsection{Instance boundaries  and bridges}
\label{sec:boundaries}

Here we will focus on those   Mastodon users that are involved in inter-instance links
and on how they can be regarded and scored as local bridges in the Mastodon user network (\textbf{Q3}). 

\paragraph{\bf Shell nodes and inter-instance edges.\ }
In Table~\ref{tab:subgraphs}, we notice that the sum of the number of nodes and  edges over the top-5 instances  are 
652\,732 and 7\,920\,818, respectively, which differs from the size of the merged network. 
 
We inspected our data looking for the reasons of the above fact, 
and found that the additional  4\,980   nodes and 306\,735   edges in the merged network take a specific role therein.  
 Given  a merged network $G_{\mathcal{M}} = \langle V, E\rangle$, we define the aforementioned two types of entities     as follows:
\begin{itemize}
\item \textit{Shell nodes}: a node $(v,i) \in V$ is said a shell node if    $\nexists (u,j) \in V: i = j \wedge (((u,j),(v,i)) \in E \vee ((v,i),(u,j)) \in E)$.
\item \textit{Inter-instance edges}:   $((u,j),(v,i)) \in E$ is said an  inter-instance edge if  $i \neq j$.
\end{itemize}

\begin{figure}[t!]
\centering
\includegraphics[width=0.7\textwidth]{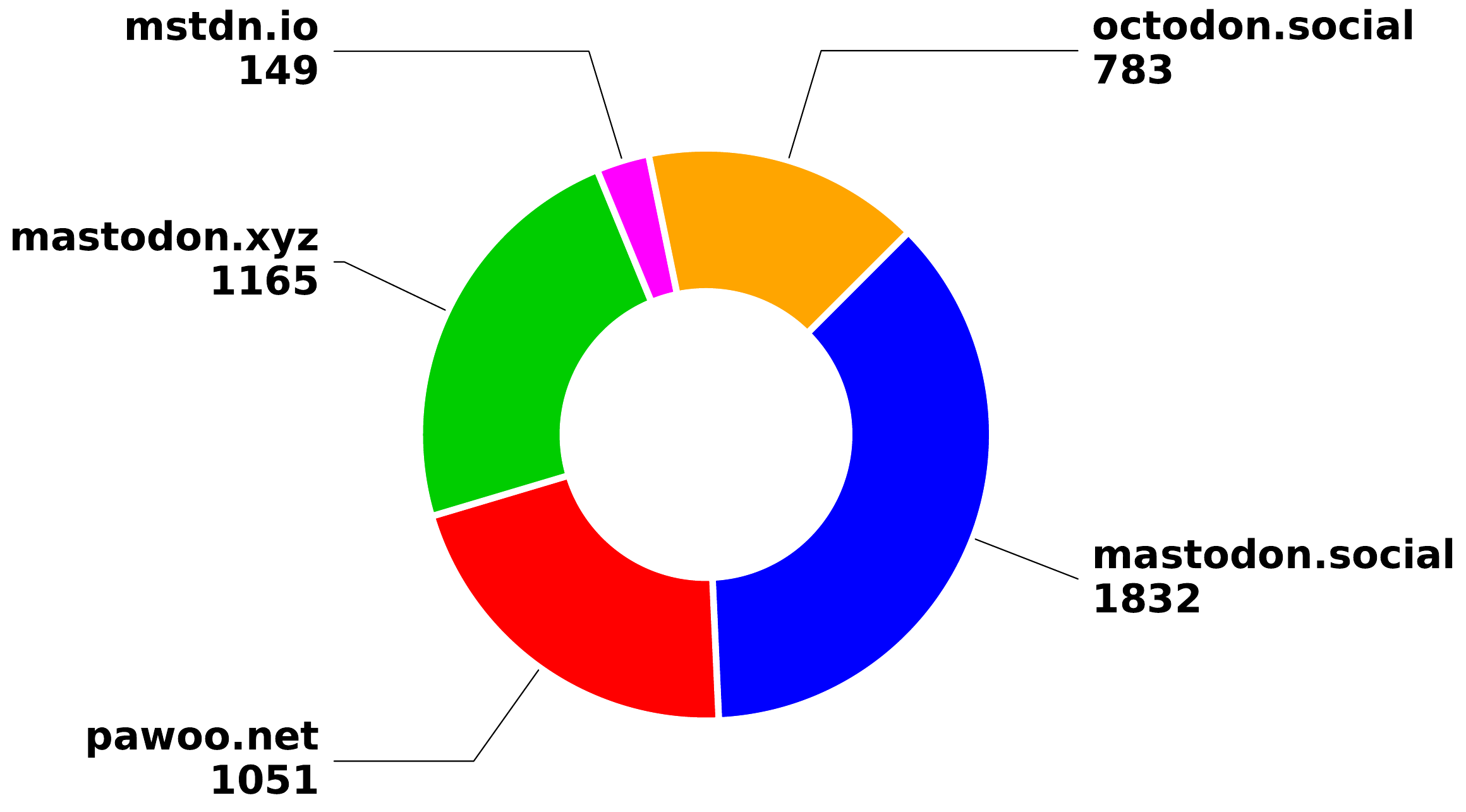}
\caption{Distribution of shell nodes within the top-5 instances composing the \textit{merged network}.}
\label{fig:shell}
\end{figure}

\noindent 
Loosely speaking, a shell node is a user linked to users of other instances only, while an inter-instance edge is a link for users of different instances.  
The Mastodon top-5 instance merged network has indeed 
4\,980   shell nodes and 306\,735  inter-instance edges;   the distribution of such nodes w.r.t. the various top-5 instances is shown in Figure~\ref{fig:shell}, whereas details about the distributions of the inter-instance edges will be considered later (cf.  Table~\ref{tab:lurker-edges}).

\begin{sidewaysfigure}
\centering
\includegraphics[width=0.95\textwidth]{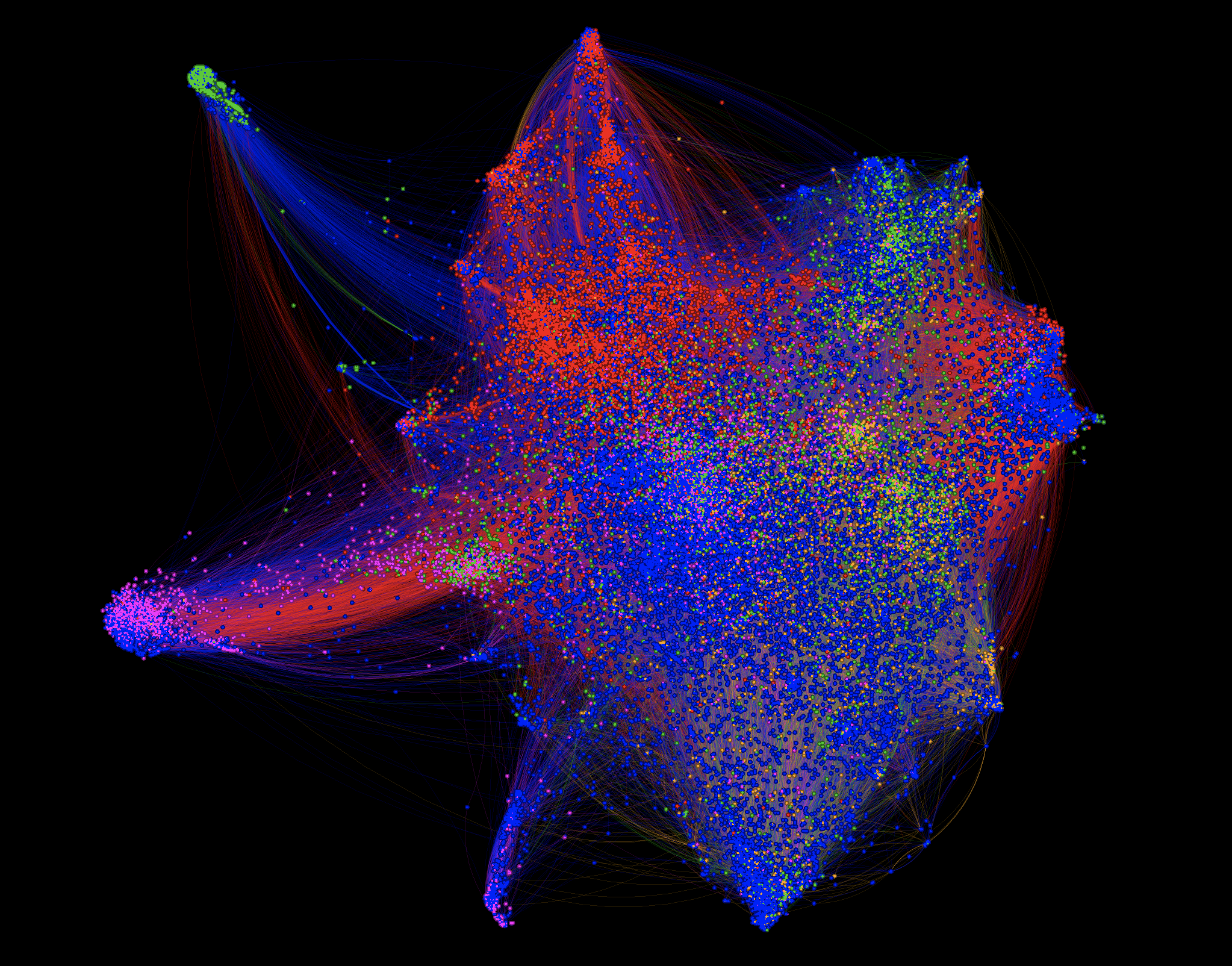}
\caption{Illustration of the inter-instance subnetwork. Nodes correspond to users belonging to the instances composing the top-5 instance \textit{merged network} and only inter-instance edges are drawn. Nodes are colored according to their membership instances, i.e., \textsl{mastodon.social} (\textit{blue}), \textsl{pawoo.net} (\textit{red}), \textsl{mastodon.xyz} (\textit{green}), \textsl{mstdn.io} (\textit{magenta}), and \textsl{octodon.social} (\textit{orange}). The color of an edge corresponds to the color of the source instance. The displayed layout is based on the force-directed drawing \textit{ForceAtlas2} model. \textit{(Produced by using the Graphistry service, available at https://www.graphistry.com.)}}
\label{fig:visualization}
\end{sidewaysfigure}

\begin{figure}[p!]
\centering
\begin{tabular}{cc}
\includegraphics[width=0.47\textwidth, height=0.25\textheight]{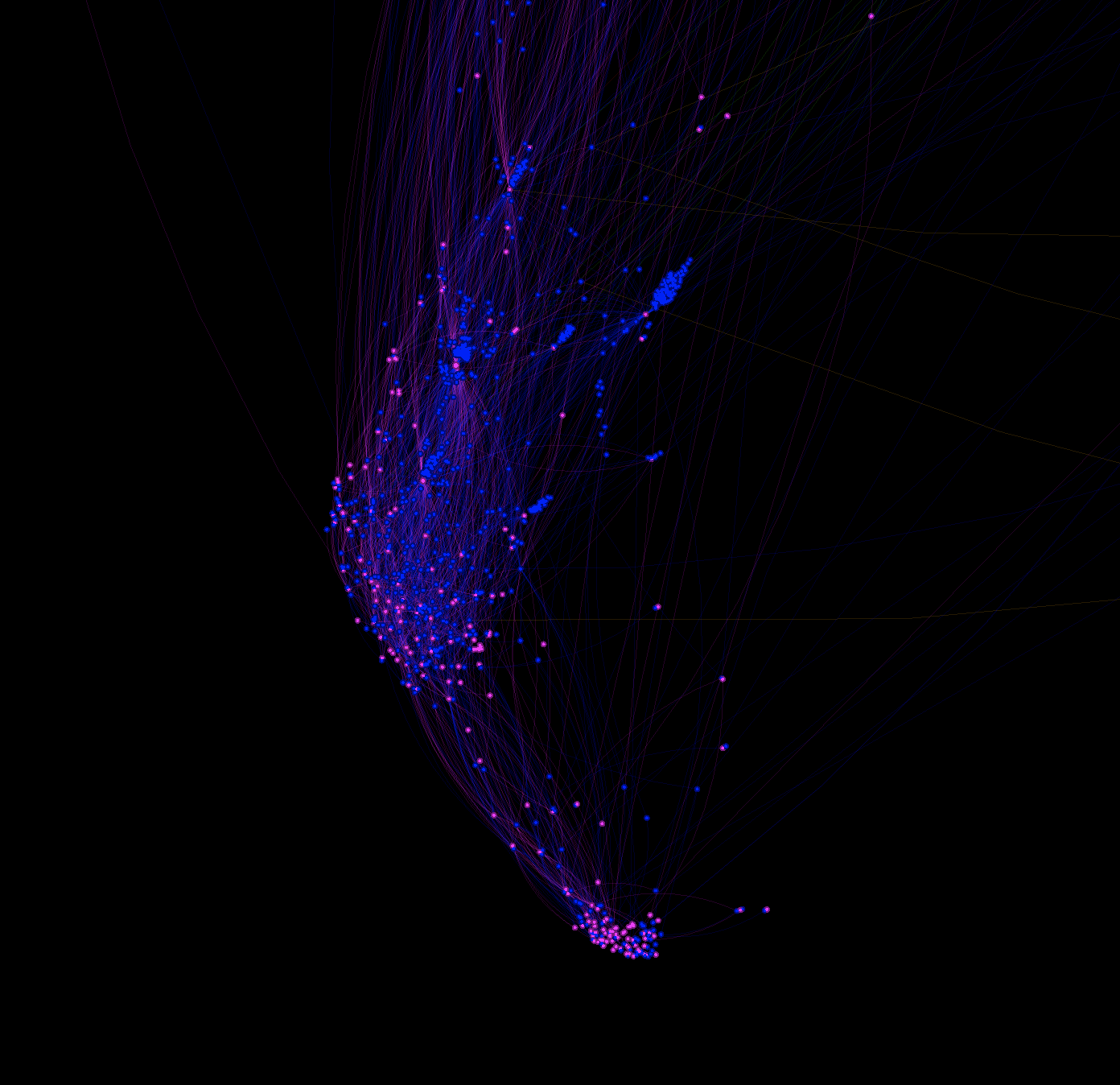} &
\includegraphics[width=0.47\textwidth, height=0.25\textheight]{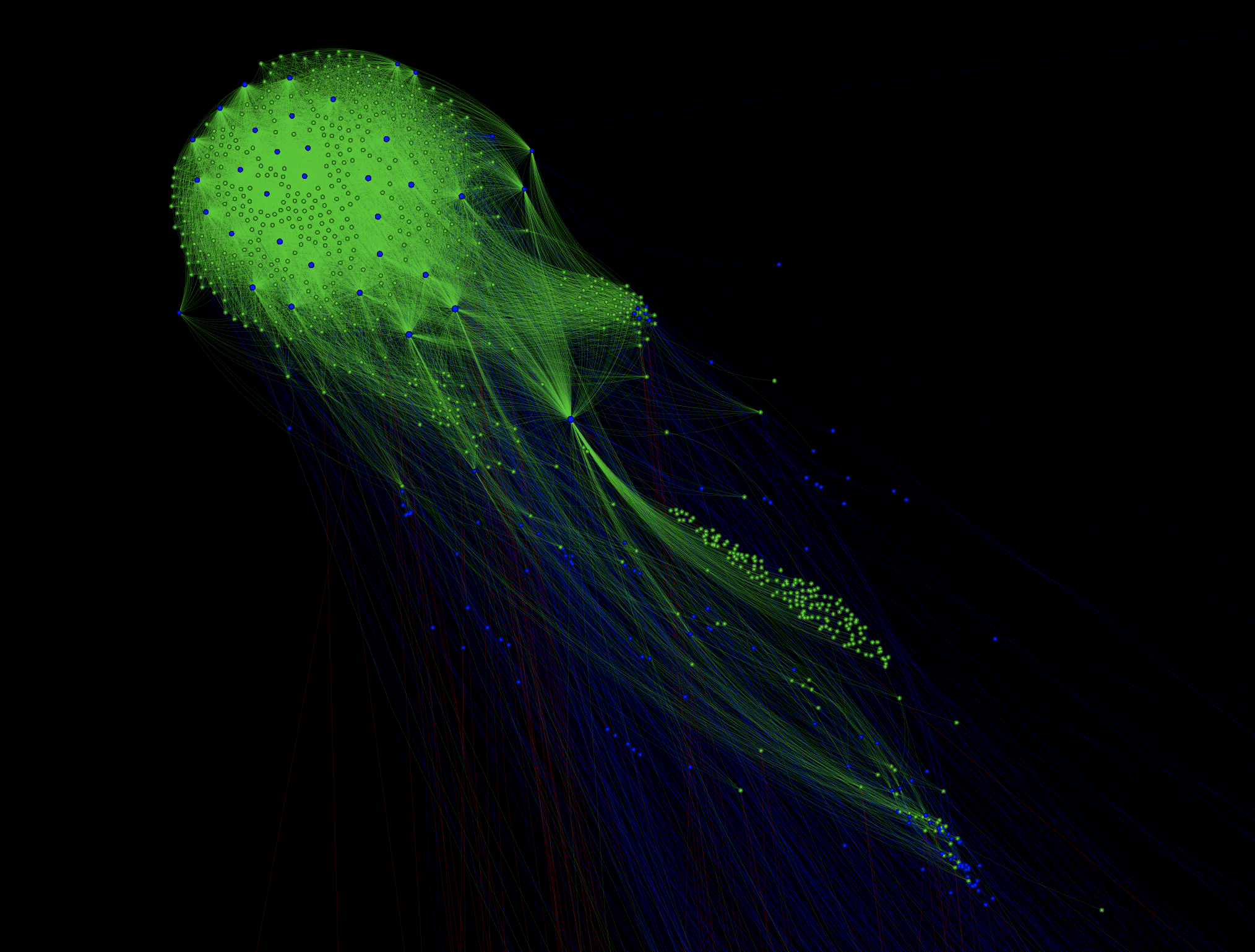} \\
(a) & (b) \\
\includegraphics[width=0.47\textwidth, height=0.25\textheight]{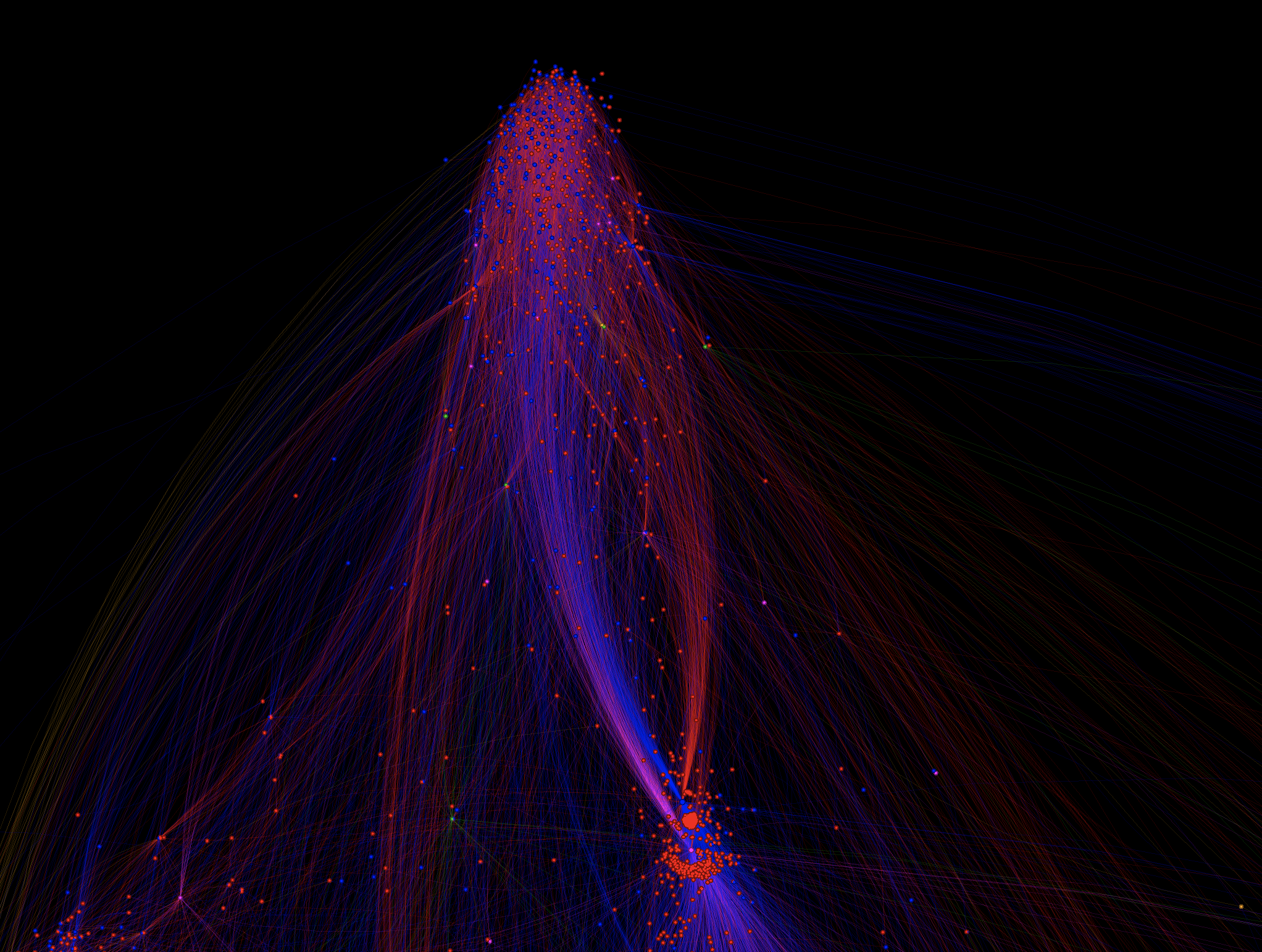} &
\includegraphics[width=0.47\textwidth, height=0.25\textheight]{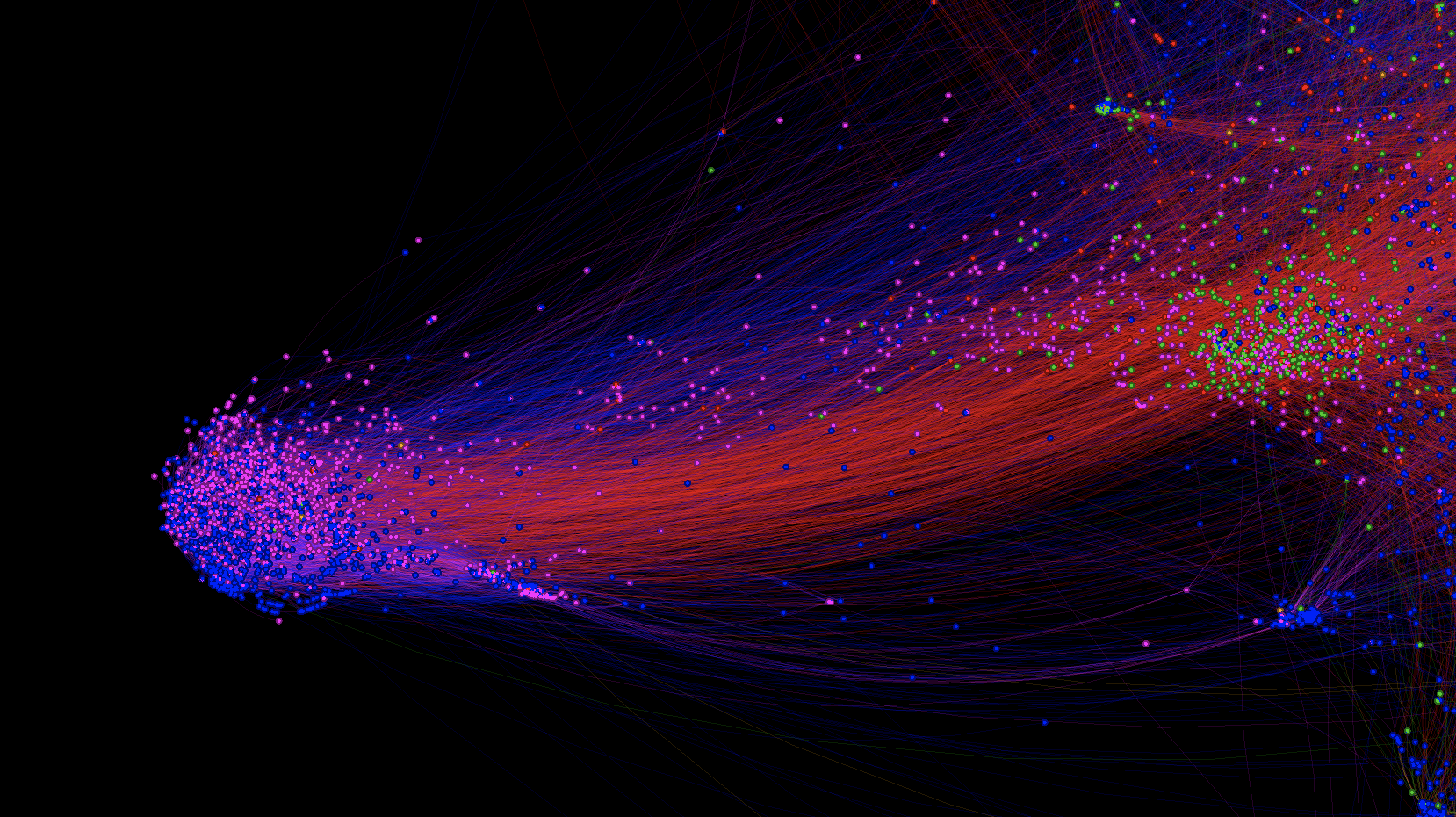} \\
(c) & (d) \\
\includegraphics[width=0.47\textwidth, height=0.25\textheight]{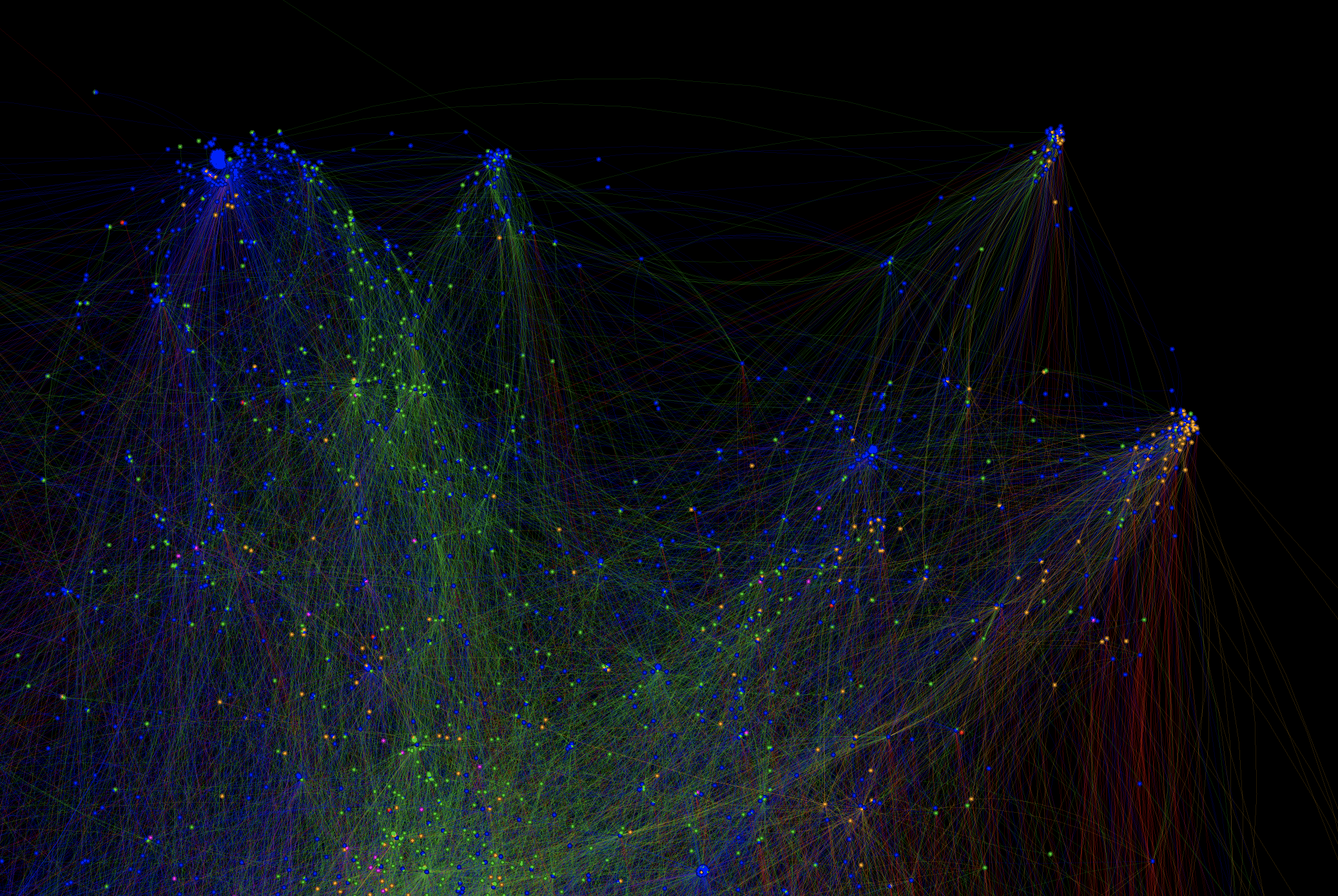} &
\includegraphics[width=0.47\textwidth, height=0.25\textheight]{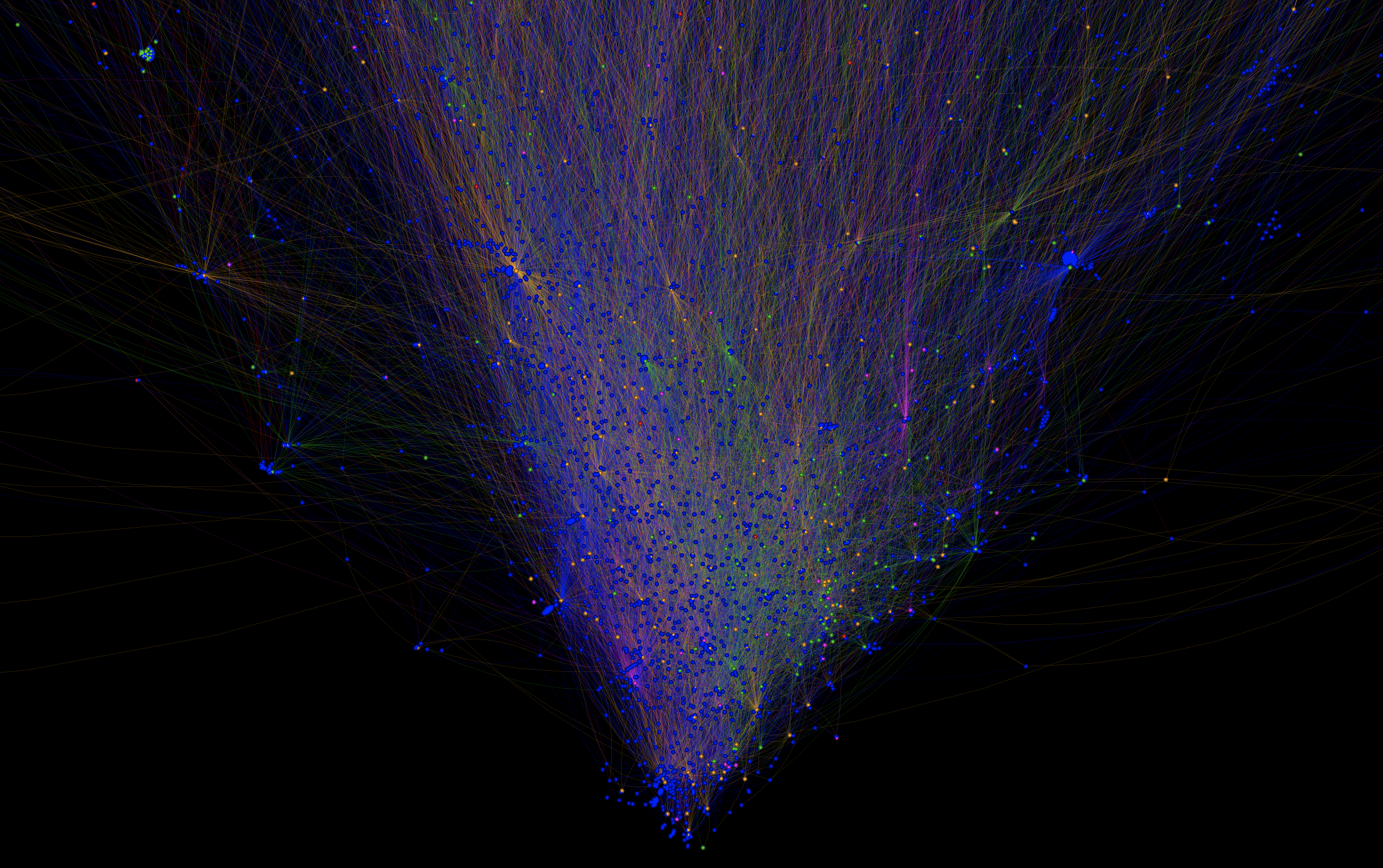} \\
(e) & (f) \\
\end{tabular}
\caption{Zoom-in views from Fig.~\ref{fig:visualization}. Nodes are colored according to their membership instances, i.e., \textsl{mastodon.social} (\textit{blue}), \textsl{pawoo.net} (\textit{red}), \textsl{mastodon.xyz} (\textit{green}), \textsl{mstdn.io} (\textit{magenta}), and \textsl{octodon.social} (\textit{orange}).   The color of an edge corresponds to the color of the source instance.  }
\label{fig:details}
\end{figure}

\paragraph{\bf Visualization of the inter-instance subnetwork.\ }
In order to get more insights into the linkage between instances, we visually inspected the inter-instance subnetwork, which models the edges connecting nodes that belong  to different instances in the top-5 instance merged network.  

As it can be observed from Figure~\ref{fig:visualization}, several interesting patterns emerge. The first eye-catching aspect  is the pervasiveness of \textsl{mastodon.social} (colored in \textit{blue}), which appears to be dominant in establishing user relationships across instances. The roots of this phenomenon plausibly lie in the relevance of \textsl{mastodon.social} since it is commonly recognized as one of the supporting pillars of the Fediverse and the first instance of the Mastodon project. The central area of Figure~\ref{fig:visualization} is characterized by  a particularly intense mix of colors, indicating the presence of user connections  that involve all the instances in the network; this is clearly consistent    with the key principle  of the Fediverse as an ecosystem made of independent yet cooperating instances. 

Through a detailed inspection of the network in  Figure~\ref{fig:visualization}, we also spotted some regions showing further relevant patterns, such as a strong linkage between users of some pairs of instances or a dense interleaving among multiple instances' users.  
In this regard, as shown in Figure~\ref{fig:details} (a), resp. Figure~\ref{fig:details} (b), there is a tight  connectivity among users of  \textsl{mastodon.social} with users of \textsl{mstdn.io}, resp. users of  \textsl{mastodon.social} with users of \textsl{mastodon.xyz}. 
An analogous situation can be observed in Figure~\ref{fig:details} (c), where a strong coupling  between users of \textsl{mastodon.social} and users of \textsl{pawoo.net} emerges, with the addition  of some sporadic users belonging to  \textsl{mstdn.io}, which are nonetheless well connected with the other two instances. 

Moreover, it is worth noticing how the pairwise interactions between (users belonging to) different instances occur with remarkable intensity even among the largest instances, as shown in Figure~\ref{fig:details} (c). This trait is particularly interesting, since although such instances can definitely be regarded as self-sufficient and represent stand-alone social platforms --- given their remarkable size ---    their users tend to interact outside the boundaries so as to gain a more global user   behavioral experience.  
 
 Figure~\ref{fig:details} (d) illustrates two regions of the network characterized by two different patterns: a particularly marked connection between users of \textsl{masto}- \textsl{don.social} and users of  \textsl{mstdn.io}  (on the left), which is also massively involved in a linkage with users from  \textsl{pawoo.net}, and another group of users from different instances (as shown by a mix of colors, on the right). The latter hints   at a comprehensive connectivity between all the instances composing the  merged network, as also confirmed by the identification of other regions characterized  by users  belonging to different instances, i.e.,   Figures~\ref{fig:details}~(e) and (f).  These events observed in the  merged network  reveal that the boundary spanning  spotted so far is not limited to pairwise instance links, but it involves multiple instances. As a consequence, we can argue that Mastodon users fully exploit the potential of seamless  interaction between independent instances provided by the platform. 
 
It should be noticed that the force-directed layout (\textit{ForceAtlas2})  we used for drawing the network emphasizes the peripheral positioning of     portions of the network, such as those corresponding to the above cases,  in which there exists a   higher connectivity among a bunch of  nodes of two or few  instances than with nodes of the other instances.

\paragraph{\bf Bridges.\ }
The above discussed boundary entities relate to another aspect of interest to our analysis of   the merged network, which is the presence of nodes connected by edges   acting as local \textit{bridges}  at varying degrees.

An effective method   to identify such  edges is to measure for each pair of linked nodes their  \textit{topological overlap}, or normalized embeddedness~\cite{Onnela2007},  which  
is the fraction of common neighbors a pair of connected vertices has. 
Indeed, edges acting as bridges are expected to share few or no neighbors, and in fact the topological overlap     enables smoothing the notion of local bridge, so that the lower the topological overlap  of a linked pair of nodes, the higher the strength of their link as local bridge. 
Originally conceived  for undirected networks, the topological overlap has also been adapted to directed networks. 
 Following~\cite{Tagarelli14},  the \textit{directed topological overlap} ($DTO$) for an edge $(u, v)$ is defined as:
$$
DTO(u, v) = \dfrac{|\mathcal{N}^{out}_{u} \cap  \mathcal{N}^{in}_{v}|}{(|\mathcal{N}^{out}_{u}| - 1) + (|\mathcal{N}^{in}_{v}| - 1) - |\mathcal{N}^{out}_{u} \cap  \mathcal{N}^{in}_{v}|}
$$
where $\mathcal{N}^{out}_{u}$ and $\mathcal{N}^{in}_{v}$ denote the out-neighbors and in-neighbors of the nodes $u$ and $v$, respectively.  
Note that the $DTO$ is  defined only for edges $(u, v)$ such that $|\mathcal{N}^{out}_{u}| > 1$ and/or  $|\mathcal{N}^{in}_{v}|> 1$; otherwise, for isolated dyads,  the $DTO$ is assumed to be zero.

Since our focus   is on users rather than links, we define  the \textit{node-centric DTO} ($nDTO$) of a   node $u$ as follows:
$$
nDTO(u) = \frac{1}{| \mathcal{N}^{in}_{u} \cup \mathcal{N}^{out}_{u} |} \left( \sum_{v \in \mathcal{N}^{in}_{u}} DTO(v,u) + \sum_{v \in \mathcal{N}^{out}_{u}} DTO(u,v) \right)
$$ 

 The application of the node-centric $DTO$ measure  produces a ranking of the nodes, whereby higher ranks (i.e., lower scores) correspond to stronger  bridge nodes. 
 (Note that, for the sake of readability, in the $DTO$ and $nDTO$ definitions we have omitted  the reference to the membership instances of $u$ and $v$).

Table~\ref{tab:bridges} shows the percentage of nodes corresponding to selected percentiles of $nDTO$ score over each of the top-5 instances as well as the whole merged network. 
As it can be noted already for the 5-$th$ percentile, it   stands out that  more than a half of the node set is   identified as bridges, with peaks above 75\% in \textsl{mastodon.xyz} and  \textsl{mstdn.io}.

We also quantified the existence of nodes showing a strong status as bridges. We identified   such nodes, dubbed  \textit{strong-bridges}, as the nodes having a $nDTO$ score equal to zero. 
 In Table~\ref{tab:bridges}, we report the number of strong bridges  for each of the top-5 instances: compared to the total number of nodes, the  percentage of  strong-bridges appears to be always  significant, ranging from   about 20\% (\textsl{octodon.social}) to about 78\% (\textsl{mastodon.xyz}).

We also evaluated the impact of source and sink nodes on the overall percentage of strong bridges found in our analyzed networks. As reported in Table~\ref{tab:bridges}, even by filtering out such nodes, the portion of strong bridges remains evident, unveiling a relatively low bias due to   source and sink nodes at least in the largest networks, where the percentage of  strong bridges ranges from about 13\% in \textsl{pawoo.net} to above 27\% in \textsl{mastodon.social}.

\begin{table}[t!]
\centering
\caption{Percentage of Mastodon users regarded as strong-bridges and  bridges for  selected cut-off thresholds of  the  $nDTO$ score percentiles.}
\label{tab:bridges}
\scalebox{0.9}{
\begin{tabular}{|l||c|c||c|c|c|}
\hline
\multirow{2}{*}{network} & \multirow{2}{*}{\#nodes} &  {\#{strong-bridges} (\%)} & \multicolumn{3}{c|}{{$nDTO$ score}}\\
\cline{4-6}
  &   &  {\small \textit{(w/o sources and sinks)}} &  5$th$ &  10$th$ & 25$th$ \\
\hline \hline

\multirow{2}{*}{\textsl{mastodon.social}} & \multirow{2}{*}{305\,968} & 121\,585  (39.7\%) & \multirow{2}{*}{61.5\%} & \multirow{2}{*}{64.6\%} & \multirow{2}{*}{70.9\%} \\  
& & 81\,112 (26.5\%) & & & \\ \hline

\multirow{2}{*}{\textsl{pawoo.net}} & \multirow{2}{*}{306\,753} & 97\,590 (31.8\%) & \multirow{2}{*}{50.6\%} & \multirow{2}{*}{53.9\%} & \multirow{2}{*}{61.8\%} \\ 
& & 38\,842 (12.7\%) & & & \\ \hline

\multirow{2}{*}{\textsl{mastodon.xyz}} & \multirow{2}{*}{16\,076} & 12\,521 (77.9\%) & \multirow{2}{*}{85.3\%} & \multirow{2}{*}{86.2\%} & \multirow{2}{*}{88.8\%} \\
& & 1\,086 (6.8\%) & & & \\ \hline

\multirow{2}{*}{\textsl{mstdn.io}} & \multirow{2}{*}{16\,853} & 4\,414 (26.2\%) & \multirow{2}{*}{75.1\%} & \multirow{2}{*}{82.6\%} & \multirow{2}{*}{86.2\%} \\
& & 290 (1.7\%) & & & \\ \hline

\multirow{2}{*}{\textsl{octodon.social}} & \multirow{2}{*}{7\,082} & 1\,436 (20.3\%) & \multirow{2}{*}{64.8\%} & \multirow{2}{*}{68.9\%} & \multirow{2}{*}{74.4\%} \\
& & 626 (8.8\%) & & & \\

\hline \hline
\multirow{2}{*}{\textit{merged network}} & \multirow{2}{*}{657\,712} & 238\,042 (36.2\%) & \multirow{2}{*}{56.0\%} & \multirow{2}{*}{59.7\%} & \multirow{2}{*}{66.8\%} \\
& & 122\,927 (18.7\%) & & & \\
\hline
\end{tabular}
}
\end{table}

\subsection{Over-consumption}
\label{sec:overconsumption}
In this section, we answer our fourth research question (\textbf{Q4})  regarding  the identification of users that tend to over-consume information produced by others. 
 To this purpose, we take a particular perspective on this problem, which relies on 
 the theory of \textit{lurking behavior analysis}~\cite{Sun+14,Edelmann13}.

A key concept in this theory is that (online) social networks are characterized by a \textit{participation inequality} principle, whereby  the crowd of a social network does not actively contribute, rather it mostly  remains hidden or ``silent'', without  taking an active role in the visible participation and interactions with other members. 
This kind of users should not be  trivially regarded as totally inactive users (i.e., registered users who do not use their account to join the online community), rather   a silent user can be perceived as someone who gains benefit from  information produced by others (e.g., reading posts and comments, watching videos, etc.)    without mostly giving back to the online community; within this view, these users are also called \textit{lurkers}. 
 It has been shown in several works (e.g.,~\cite{Sun+14,Edelmann13,SorokaR06,GongLZ15,Wang0LM19}) that  lurking is normal and also an active, participative and valuable form of online behavior, including a form of cognitive apprenticeship that corresponds to  legitimate peripheral participation. In this respect, lurkers might have a great potential in terms of social capital, since they acquire knowledge from the community; therefore, when engaged, they become beneficial for the propaganda and development of the community.

 Modeling and analyzing lurking behaviors has been formulated as a eigen\-vector-centrality-based node ranking problem, which is totally content-agnos\-tic, as it does not require other information than the graph topology~\cite{Tagarelli14}. 
 The \textsf{LurkerRank} method was designed to assign each user a score expressing her/his lurking status. In \textit{\textbf{Appendix}},   we report the mathematical details of this method. 
 It should be noted that the \textsf{LurkerRank} method applies to a network graph with \textit{reversed edge-orientation}, therefore hereinafter we shall consider any edge $(u,v)$ as a link from $u$ to $v$ where $v$ is a follower of $u$.

 \begin{table}[t!]
\centering
\caption{Percentage of Mastodon users regarded as lurkers w.r.t. selected cut-off thresholds based on the \textsf{LurkerRank} score percentiles.}
\label{tab:percentiles}
\begin{tabular}{|l||c|c|c|}
\hline
\multirow{2}{*}{network} & \multicolumn{3}{c|}{$LR$} \\
\cline{2-4}
&  95$th$ & 90$th$ &  75$th$ \\
\hline 	\hline
\textsl{mastodon.social} & 2.8\% & 7.7\% & 30.1\% \\ 
\textsl{pawoo.net} & 4.3\% & 9.3\% & 28.3\% \\ 
\textsl{mastodon.xyz} & 2.0\% & 6.0\% & 73.6\% \\ 
\textsl{mstdn.io} & 4.6\% & 8.9\% & 77.6\% \\ 
\textsl{octodon.social} & 38.7\% & 41.6\% & 56.3\% \\
\hline \hline 
\textit{merged network} & 3.7\% & 9.0\% & 24.3\% \\ 
\hline
\end{tabular}
\end{table}

To answer the research question \textbf{Q4}, our main goal is to  understand whether and to what extent lurkers of an instance are  target nodes of an information flow coming either from the same instance or from a different instance.  
To this purpose, we compute the \textsf{LurkerRank} method to  each of the top-5 instance networks as well as to the   merged network. 
In Table~\ref{tab:percentiles}, we report the percentage of users  identified as lurkers for selected percentiles of $LR$ values, where $LR$ symbol is used to denote the scoring function of \textsf{LurkerRank} (cf. \textit{\textbf{Appendix}}). 
Looking at the table, we notice that the   merged network as well as each of the top-5 instances, but \textsl{octodon.social}, show  a percentage of lurkers that is below 5\% and 10\% for the 95$th$ and the 90$th$ percentile, respectively, while for \textsl{octodon.social}, the percentage values at 95$th$ and 90$th$ percentiles are comparable and set around 40\%. 
However, when extending to the 75$th$ percentile, the percentage of users increases   to at least approximately 30\% (for \textsl{mastodon.social} and \textsl{pawoo.net}), with a peak above 70\% in \textsl{mastodon.xyz} and \textsl{mstdn.io}. 
Note also that  the increment on \textsl{octodon.social} appears to be at a significantly lower rate than for the other instance networks.

\paragraph{\bf Information consumption.\ }
Once the lurkers at varying degrees  were identified within the merged network, we investigated  the links towards lurker nodes of a specific instance w.r.t. the overall incoming links, in order to understand   how much  the information flow is ``consumed'' by (i.e., it is directed to) lurkers,  and whether this occurs  internally or externally to their membership instance.

\begin{table}[t!]
\centering
\caption{Percentage of outgoing edges,  resp. incoming edges, between pairs of selected instances that correspond to edges towards, resp. from, lurkers. 
Percentiles refer to \textsf{LurkerRank} scores.}
\label{tab:lurker-edges}
\scalebox{0.77}{
\begin{tabular}{|l|l||c||c|c|c||c|c|c|}
\hline
source & target & \multirow{2}{*}{\#edges} & \multicolumn{3}{c||}{edges \textit{to lurkers}} & \multicolumn{3}{c|}{edges \textit{from lurkers}} \\
\cline{4-9}
instance & instance & & \textit{95th} & \textit{90th} & \textit{75th} & \textit{95th} & \textit{90th} & \textit{75th} \\
\hline 	\hline

\multirow{5}{*}{\textsl{mastodon.social}}  & \textsl{mastodon.social} & 3\,408\,327 & 12.6\% & 15.1\% & 25.3\% & 0.4\% & 0.6\% & 2.2\% \\
& \textsl{pawoo.net} & 45\,713 & 3.4\% & 8.3\% & 19.5\% & 1.5\% & 1.9\% & 5.3\% \\
& \textsl{mastodon.xyz} & 46\,069 & 5.8\% & 7.7\% & 44.4\% & 0.8\% & 1.0\% & 2.4\% \\
& \textsl{mstdn.io} & 29\,425 & 7.0\% & 9.6\% & 26.8\% & 0.8\% & 1.0\% & 2.4\% \\
& \textsl{octodon.social} & 29\,935 & 5.9\% & 8.7\% & 22.1\% & 0.9\% & 1.1\% & 3.0\% \\
\hline \hline

\multirow{5}{*}{\textsl{pawoo.net}}  & \textsl{mastodon.social} & 29\,572 & 48.5\% & 49.4\% & 54.7\% & 1.0\% & 1.4\% & 3.8\% \\
& \textsl{pawoo.net} & 4\,329\,562 & 13.4\% & 16.8\% & 31.8\% & 0.1\% & 0.2\% & 0.8\% \\
& \textsl{mastodon.xyz} & 944 & 0.7\% & 12.0\% & 34.4\% & 0.1\% & 0.2\% & 1.6\% \\
& \textsl{mstdn.io} & 3\,192 & 3.3\% & 6.4\% & 43.7\% & 0.5\% & 0.7\% & 2.3\% \\
& \textsl{octodon.social} & 945 & 11.7\% & 19.2\% & 39.8\% & 0.1\% & 0.1\% & 0.6\% \\
\hline \hline

\multirow{5}{*}{\textsl{mastodon.xyz}}  & \textsl{mastodon.social} & 36\,328 & 19.5\% & 21.8\% & 34.1\% & 1.0\% & 1.9\% & 10.8\% \\
& \textsl{pawoo.net} & 6\,684 & 1.3\% & 6.6\% & 17.8\% & 2.8\% & 5.9\% & 30.5\% \\
& \textsl{mastodon.xyz} & 35\,631 & 4.5\% & 8.9\% & 46.2\% & 0.0\% & 0.1\% & 3.0\% \\
& \textsl{mstdn.io} & 1\,417 & 9.5\% & 12.8\% & 41.6\% & 0.4\% & 0.6\% & 4.2\% \\
& \textsl{octodon.social} & 2\,404 & 3.2\% & 5.9\% & 14.6\% & 0.2\% & 0.6\% & 5.2\% \\
\hline \hline

\multirow{5}{*}{\textsl{mstdn.io}}  & \textsl{mastodon.social} & 28\,523 & 15.2\% & 16.7\% & 23.2\% & 1.8\% & 2.7\% & 12.6\% \\
& \textsl{pawoo.net} & 6\,691 & 1.6\% & 4.9\% & 13.3\% & 5.7\% & 8.3\% & 35.5\% \\
& \textsl{mastodon.xyz} & 803 & 6.2\% & 8.1\% & 24.7\% & 1.0\% & 1.9\% & 15.8\% \\
& \textsl{mstdn.io} & 112\,805 & 4.5\% & 6.6\% & 25.3\% & 0.0\% & 0.0\% & 0.7\% \\
& \textsl{octodon.social} & 626 & 7.7\% & 11.8\% & 22.0\% & 1.3\% & 2.7\% & 18.4\% \\
\hline \hline

\multirow{5}{*}{\textsl{octodon.social}}  & \textsl{mastodon.social} & 34\,158 & 17.6\% & 19.1\% & 28.0\% & 2.5\% & 7.5\% & 14.5\% \\
& \textsl{pawoo.net} & 84 & 0.0\% & 0.0\% & 0.0\% & 4.8\% & 4.8\% & 6.0\% \\
& \textsl{mastodon.xyz} & 2\,281 & 4.4\% & 5.9\% & 18.8\% & 1.1\% & 3.0\% & 5.8\% \\
& \textsl{mstdn.io} & 941 & 10.1\% & 13.3\% & 39.3\% & 1.0\% & 1.9\% & 5.5\% \\
& \textsl{octodon.social} & 34\,493 & 20.8\% & 24.3\% & 37.2\% & 0.5\% & 1.2\% & 7.7\% \\

\hline
\end{tabular}
}
\end{table}

We report the results of our analysis in Table~\ref{tab:lurker-edges} under the column ``edges \textit{to lurkers}'', for each pair of the top-5 instances --- including self-pairing, i.e., within-instance links ---     and for various lurking score percentiles. 
At a first glance, it can be noted a certain variety in the percentage values, which indicates  a remarkable differentiation of information  consumption  by lurkers within and across the various instances. 

 On the one hand, there is an evidence of information flow directed to lurkers from inside their membership instance, although this happens at different extents; in particular,  at the 95$th$ percentile, the percentage of links directed to lurkers ranges from 4.5\% in \textsl{mstdn.io} and \textsl{mastodon.xyz} to about 21\% in  \textsl{octodon.social}. 
 
 On the other hand, however, there is also a remarkable amount  of information flow directed to lurkers from outside their membership instance. 
 In this respect,  the \textsl{mastodon.social} instance turns out to be   the best target for lurkers, given the highest percentages of links coming from the other instances (and  \textsl{mastodon.social} itself) and directed to lurkers. In particular,  we notice  a considerable amount of information flow from  \textsl{pawoo.net} to    lurkers in \textsl{mastodon.social}, which is  about 50\% of the connections from \textsl{pawoo.net} to    \textsl{mastodon.social}.  
   Also, \textsl{mastodon.xyz} and \textsl{mstdn.io} lurkers tend to absorb most information   from  outside, while  \textsl{pawoo.net} is particularly  relevant  as information producer for its own users as well as for users of the other instances. 
   
\begin{figure}[t!]
\centering
\includegraphics[width=0.6\textwidth]{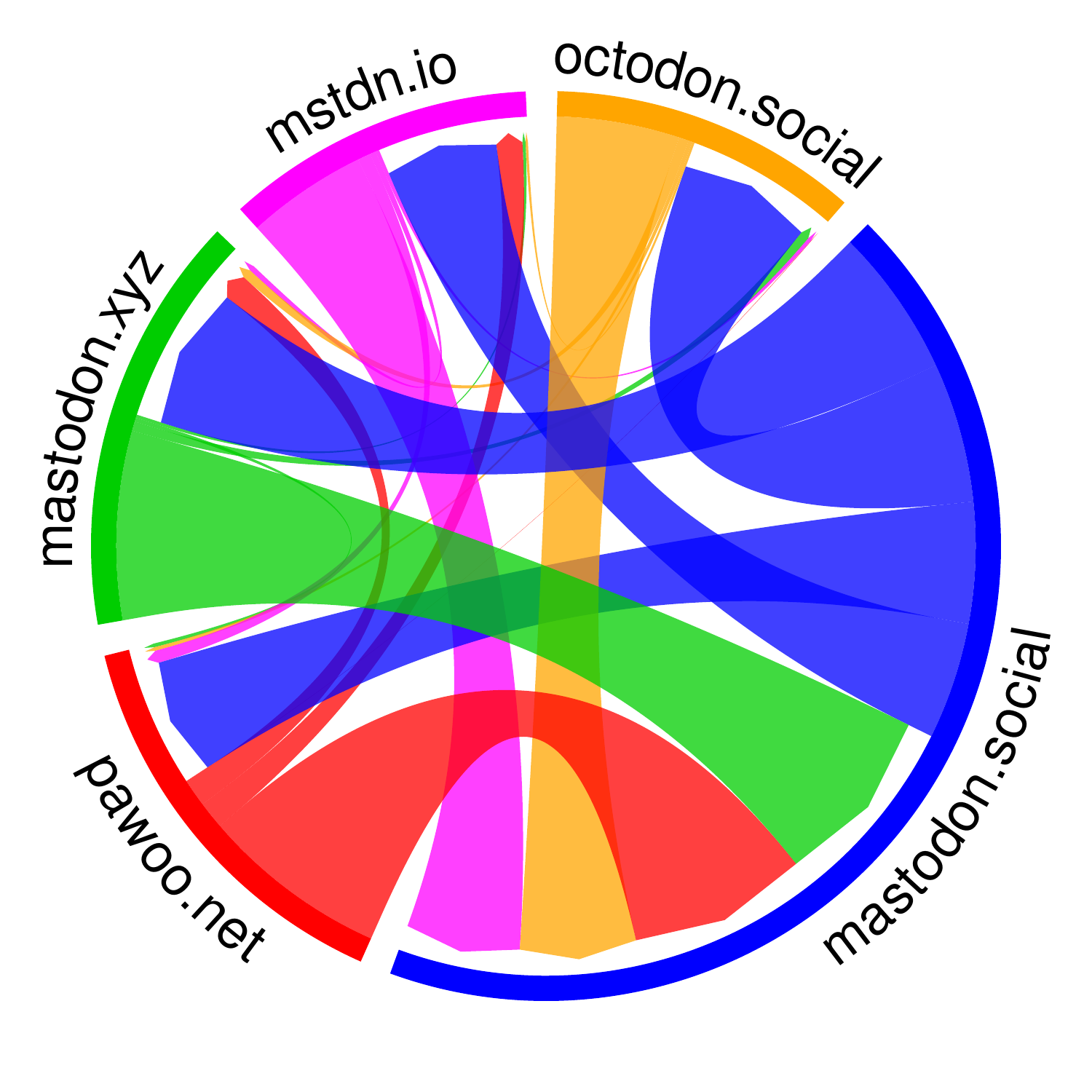}
\caption{Graphical illustration of the information flow between the top-5 Mastodon instances, modeled from information producers to information consumers. Self-loops and flow values (cf. Table~\ref{tab:lurker-edges}) are omitted to avoid cluttering.  Each flow has the same color as the source instance.}
\label{fig:flow}
\end{figure}

   Overall, the above remarks highlight an important trait of the Mastodon network as a mix of within-instance and across-instance information consumption of its users. 
Figure~\ref{fig:flow} provides an illustration of the information flow between the top-5 instances, which complements our understanding from the results shown in  Table~\ref{tab:lurker-edges}  by   highlighting a sort of mutual reinforcement among such instances, in terms of information production, resp.  consumption, behaviors exhibited by their users.

\paragraph{\bf Information spreading.\ }
Analogously to the previous analysis,  
we examined how the outgoing links from an instance might be originated by  lurkers.

Clearly, as  expected from the definition of lurking behavior,  lurkers do not    contribute much to the diffusion of the information they consume, which is  indicated by the small percentage values reported in Table~\ref{tab:lurker-edges}, under the column ``edges \textit{from lurkers}''.
However, some exceptions stand out. In particular, already at the 95$th$ percentile,   lurkers of \textsl{mastodon.xyz}, \textsl{mstdn.io} and \textsl{octodon.social} contribute to spread information  towards \textsl{pawoo.net} in a non-negligible way. 
This trend strengthens at  the 90$th$ percentile and, for \textsl{mastodon.xyz} and \textsl{mstdn.io},   is boosted at  the 75$th$ percentile  (with peaks above 35\%  from \textsl{mstdn.io} to \textsl{pawoo.net}). 
 The above  is remarkable  as we recall that   \textsl{pawoo.net} tends to attract lurkers belonging to other instances (e.g., \textsl{mastodon.social} and \textsl{octodon.social}), and conversely, we also spotted that lurkers from  all the  other instances (especially from  \textsl{mastodon.xyz}, \textsl{mstdn.io}, \textsl{octodon.social}) might contribute to the diffusion of information towards \textsl{pawoo.net}.

\subsection{Dual role users}
Our   fifth question (\textbf{Q5}) concerns unveiling the existence of Mastodon users that take a twofold role as lurkers and bridges.   
To this purpose, for each of the top-5 instances and the   merged network, we analyzed the overlap  between a set of lurkers and a set of bridge users,   selected from their respective ranking solutions  according to the percentile thresholds used in the previous analysis, such that each set pair  refers to the same percentile proportion (i.e., 95$th$ vs. 5$th$,  90$th$ vs. 10$th$, and 75$th$ vs. 25$th$).

\begin{table}[t!]
\centering
\caption{Dual role users in the top-5 instance networks and  in the merged network.}
\label{tab:multiple-behaviors}
\scalebox{0.9}{
\begin{tabular}{|l||c||c|c|c|}
\hline
\multirow{2}{*}{network} & \multirow{2}{*}{\#nodes} & $LR$@95$th$   & $LR$@90$th$   & $LR$@75$th$  \\
&   &   $\cap$ $nDTO$@5$th$ &  $\cap$ $nDTO$@10$th$  &   $\cap$ $nDTO$@25$th$\\
\hline \hline
\textsl{mastodon.social} & 305\,968 & 0.7\% & 5.1\% & 26.1\% \\
\textsl{pawoo.net} & 306\,753 & 1.6\% & 4.9\% & 20.4\% \\
\textsl{mastodon.xyz} & 16\,076 & 1.1\% & 4.9\% & 72.3\% \\
\textsl{mstdn.io} & 16\,853 & 0.5\% & 7.3\% & 76.6\% \\
\textsl{octodon.social} & 7\,082 & 37.6\% & 39.5\% & 51.5\% \\
\hline\hline
\textit{merged network} & 657\,712 & 1.5\% & 6.0\% & 18.9\% \\
\hline 
\end{tabular}
}
\end{table}

Table~\ref{tab:multiple-behaviors} reports on the results of our analysis.  
The  \textsl{octodon.social} instance shows by far the largest percentage of users exhibiting the dual role, which is already above 37\% w.r.t. the toughest overlap (i.e., $LR$@95$th$ $\cap$ $nDTO$@5$th$), then settling around 51\% for the smoothest overlap (i.e., $LR$@75$th$ $\cap$ $nDTO$@25$th$). 
 Note that  the latter point corresponds to an increase rate that is much lower than for the other instances, where the percentages of dual roles keep far below 10\% w.r.t. the two largest overlaps while increasing up to a minimum of 20\%  (\textsl{pawoo.net}) and a maximum of 77\%  (\textsl{mstdn.io}) w.r.t. the overlap $LR$@75$th$ $\cap$ $nDTO$@25$th$. 
Interestingly, this is in accord with the higher, resp. lower, smoothness  in the role identification shown by \textsl{octodon.social}, resp. the other instances,  for varying scoring percentiles, as we already observed in our previous analysis (cf. Table~\ref{tab:lurker-edges}). 
 Moreover, the difference in percentages corresponding to the $LR$@75$th$ $\cap$ $nDTO$@25$th$ between the two largest instances (i.e., \textsl{mastodon.social} and \textsl{pawoo.net}) and the other three instances also depends on the size of their respective user-bases: in fact, as the number of users in an instance gets smaller, the volume of information produced becomes  more limited, and  hence    their  users tend not only to consume  it but also to act        as \textit{information flow facilitators}; by doing this, they can contribute keeping the instance sustainable with fresh contents and timely interactions.

Considering the   merged network, there is evidence of a certain presence of dual role users --- thus   indicating that a dual role behavioral phenomenon can also occur   in a cross-instance context --- with   percentages  that are in line with some of the instances, particularly \textsl{pawoo.net}.

\paragraph{\bf Alternate role users.\ }
Here we consider our sixth research question (\textbf{Q6}). The goal is to  understand whether by mixing the  scales of observation, i.e., either locally within an instance or globally at the level of the merged network,  distinct   behaviors of users may arise. 
As for the previously analyzed research questions, we focus on lurkers and bridge users, thus aiming   to identify whether  users can be regarded as  lurkers inside their membership instance yet as   bridges in a cross-instance environment, and vice versa.  
 
 \begin{table}[t!]
\centering
\caption{Mastodon users who behave differently according to the observation scale, i.e., locally within their instance (denoted with superscript $(L)$)  or globally at the level of merged network (denoted with superscript $(G)$)}
\label{tab:multi-scale}
\scalebox{0.7}{
\begin{tabular}{|l||c|c|c|c|c|c|c|}
\hline
& \textsl{mastodon.social} & \textsl{pawoo.net} & \textsl{mastodon.xyz} & \textsl{mstdn.io} & \textsl{octodon.social} \\
\hline \hline
\#users & 305\,968 & 306\,753 & 16\,076 & 16\,853 & 7\,082 \\
\hline
{\small $LR$@95$th^{(L)}$ $\cap$ $nDTO$@5$th^{(G)}$} & 0.6\% & 1.6\% & 0.8\% & 0.3\% & 1.2\% \\
{\small $LR$@90$th^{(L)}$ $\cap$ $nDTO$@10$th^{(G)}$} & 4.8\% & 5.0\% & 2.7\% & 2.4\% & 36.2\% \\
{\small $LR$@75$th^{(L)}$ $\cap$ $nDTO$@25$th^{(G)}$} & 25.0\% & 21.2\% & 66.7\% & 72.7\% & 45.1\% \\
\hline
{\small $nDTO$@5$th^{(L)}$ $\cap$ $LR$@95$th^{(G)}$} & 0.2\% & 3.1\% & 0.3\% & 0.1\% & 0.1\% \\
{\small $nDTO$@10$th^{(L)}$ $\cap$ $LR$@90$th^{(G)}$}& 0.7\% & 12.2\% & 0.4\% & 0.2\% & 0.1\% \\
{\small $nDTO$@25$th^{(L)}$ $\cap$ $LR$@75$th^{(G)}$} & 13.3\% & 26.6\% & 2.7\% & 1.2\% & 0.5\% \\
\hline
\end{tabular}
}
\end{table}

In Table~\ref{tab:multi-scale}, we report the percentage of users of a given instance  that are identified as users showing a lurking role locally and a bridging role globally  (upper subtable). 
 As it can be noted, while a few cases (below 2\%) are already identified w.r.t.  $LR$@95$th^{(L)}$ $\cap$ $nDTO$@5$th^{(G)}$, this alternate behavior becomes more evident w.r.t. larger overlaps, on all instances though at different extents.  
 These results allow us to understand how the information flow moves within a decentralized context. An intra-instance (or local) lurker is a user who tends to absorb information, while an inter-instance (or global) bridge is a user who contributes to connect  multiple regions of a network of instances. It follows that the users having this dual scale-dependent role are those who, while consuming locally produced information, enable the information coming from their instances to flow into the Fediverse, thus becoming potential information facilitators.   
 Moreover, as already partially unveiled in our previous analysis on dual role users,  the higher percentages of alternate role users generally found for   instances with a smaller user base suggest a tendency of users in such instances   to act as a touch-point and interconnect different regions that cross the instance boundaries.

 In Table~\ref{tab:multi-scale}, we also report the percentage of users of a given instance  that are identified as users showing a bridging role locally and a lurking role globally  (bottom subtable).  
 We observe that the percentage values are generally much lower than the previously discussed behavioral case. This should not be surprising since  if a user takes a within-istance bridge role, s/he is already committed to broker information  and hence will likely be  less inclined to absorb information from the outside. Nonetheless, in this scenario,  \textsl{mastodon.social} and \textsl{pawoo.net}  represent  an exception,   showing non-negligible overlaps of alternate role users under less restrictive percentile thresholds. We tend to ascribe this phenomenon to   aspects related to the topology of those instances; in particular,  the sparsity of connections over a large user base would favor some users to absorb information from other instances while acting as bridges locally.

\section{Discussion}
\label{sec:discussion}

Here we summarize  the main findings that raised from our extensive analysis of the Mastodon user relations. 

To answer our first research question (\textbf{Q1}), we explored the main structural characteristics of the Mastodon user network. 
 Among the noteworthy facts, we observed a lack of degree correlation, which  should be ascribed  to a form of   spontaneous connectivity between users that relates to the absence of boosting mechanisms for ``artificial'' interactions, such as those due to the widely used recommendation strategies adopted by the centralized OSNs. 
 From a mesoscopic perspective, based on Louvain, Leiden, and Infomap community detection methods, the user networks exhibit a moderately high modularity (around 0.7) and a high number of communities; this trait, which    indicate the existence of small densely connected   groups of users tailored to specific shared interests, appears to be consistent with the spontaneous connectivity trend in Mastodon.  
 
 Remarkably, all the  specific traits  discovered on    the full  user-network remained valid also after our step of graph pruning aimed at removing irrelevant instances. 
 This  was further strengthened when  we considered  a set of instances able to represent the entire Mastodon user network, as outlined by our second research question (\textbf{Q2}). To this purpose, supported by some pertinent results from the study in~\cite{LaCava2021}, we focused on the five most relevant instances in   Mastodon. After evaluating their main structural features and finding high consistency with the results obtained on the full user-network,   we concluded that the top-5 instance    network  can be regarded as representative of the whole Mastodon user network, and indeed we used it in our subsequent tasks of user behavior analysis.

To answer our third research question (\textbf{Q3}), we investigated the linkage between Mastodon users  accounting for the instance boundaries. We indeed found out a significant fraction of inter-instance links and of shell nodes (i.e., users having connections with other instances' users only), 
thus unveiling  an evident boundary-spanning phenomenon,  as also confirmed by our visual inspection in Figs.~\ref{fig:visualization} and~\ref{fig:details}. We delved into the  boundary-spanning  mechanisms in Mastodon through the identification of  users acting as bridges at varying degrees. 
 To this purpose, by leveraging the notion of directed topological overlap, we discovered a widespread presence of bridge users, with  a non-negligible fraction of what we called  strong-bridges, i.e., users having a topological overlap equal to zero. Interestingly, this still holds  even by removing the source and sink nodes from the network. Therefore, we can state the existence of structurally strategic nodes holding connections between across-instance regions of the network, which   positively impacts on the effectiveness of information flow between the users over all Mastodon.

As for our fourth research question (\textbf{Q4}), we modeled     the information over-consumption phenomenon through the Mastodon user network in terms of lurking behaviors. As thoroughly discussed in the literature, lurkers are silent users who tend to mostly consume information from the others' actions rather than produce information; but at the same time, by holding a certain  social capital and given their pervasiveness in a social network, such users might significantly contribute to boundary spanning and information flow phenomena. 
In this regard, we built our analysis  upon a theoretically well-founded content-agnostic eigenvector-centrality ranking method,   \textsf{LurkerRank}. 
 Our  goal was twofold: to  understand whether and to what extent lurkers of an instance are  target nodes of an information flow coming from other users, and whether this involves  the membership instance or the other instances.  
 In this regard, we found out that lurkers are present, at varying degrees, over all the selected instances under study,  with \textsl{mastodon.social} being the preferred instance for  information consumption by lurkers. In general,   information consumption is not confined to the membership instance, but it extends beyond the boundaries of the instances, so as to further capitalize on the information exchanged through different regions of the Mastodon user network. 
 Furthermore, we unveiled that lurkers are also      involved in information spreading processes between instances, even in a non-negligible way as it happens for users in \textsl{pawoo.net} that are linked to (i.e., follow) lurkers of the other instances.

Our last research questions regarded the existence of users who show a dual lurker-bridge role, either  simultaneously through the whole user network and the instance-specific subnetworks (\textbf{Q5}), or     alternately as a function of the observation scale, i.e., inter-instance and intra-instance perspective (\textbf{Q6}).    
We found a relatively small   fraction of users acting both as lurkers and bridges within their own instances; since these users normally over-consume  but have also the potential of  disseminating information, they could be regarded  as information flow facilitators. This trait is present through all the merged network, and is particularly evident in the smallest instances, where the produced information is limited to the size of the audience in those instances, and hence an amplification is needed. 
Concerning the alternate and scale-dependent behavior, we spotted the existence of users acting as local (i.e., on their own instances) lurkers and global (i.e., between instances) bridges, whereas the contrary does not hold. Reasonably, the former trait allows users to disseminate information from their own instances outwards, while the latter is unnecessary as they are already responsible for the intra-instance information spreading.  
As a final remark, we believe that such dual/alternate-role users can be regarded as highly strategical ones, as their complementary structural functionality makes them ideal candidates to determine the speed and scope by which the information flows within Mastodon, and more generally, in a decentralized social context.

\section{Conclusions}
\label{sec:conclusions}

Decentralized Online Social Networks (DOSNs) aim to bring the social paradigm back to its roots made up of spontaneous interactions and genuine interests, in contrast  to the marketing-driven engagement mechanisms typically adopted by the centralized OSNs. To guarantee a user-centric vision, DOSNs support the creation of independent and self-hosted servers seamlessly connected among each other. 
Nevertheless, this   metamorphosis of the online social media environment could change how some of the fundamental components underlying  human relationships appear and evolve via the Internet.

In this work, we have provided a number of insights into DOSN user relations and behaviors,  using    as a case in point Mastodon, the most-known service of the Fediverse. 

We analyzed the Mastodon user network 
to answer six research questions encompassing the main structural characteristics of the following user relations, the impact due to the most representative instances on the user network,  across-instance boundary spanning and bridges, over-consumption and information flow, dual and alternate role users.

Our future work plan includes further investigation on the impact that decentralization has on user behaviors and how the latter adapt to allow information   flowing quickly and across instances. 
 In this regard, we are interested in defining a suitable multilayer network model for the user relations in order to support the characterization and analysis of the duality and mutual reinforcement between  information-production and information-consumption behaviors as a function of the users' timelines and   instance-based contexts.


\section*{Appendix} 
Lurker ranking methods, originally proposed in~\cite{TagarelliI13,Tagarelli14},  are designed to mine silent user behaviors in the network, and hence to associate  each user  with a score  indicating her/his lurking   status.     
Since the basic assumption of    lurking behaviors is related to the \textit{amount of information a user receives},  the key idea in the definition of lurker ranking methods  is that  the strength of  a user's lurking status can be determined based on three main  principles, namely     over-consumption, authoritativeness of the information received,  non-authoritativeness of the information   produced. 

The first principle corresponds to the evidence of a disproportion between  information-consumption over information-production exhibited by a user. 
The second principle relates to the importance as information producers of a user's followees (i.e., in-neighbors), while the third principle related to the   low importance as information producer of a user with respect to her/his followers (i.e., out-neighbors). 

These   principles are implemented in a ranking model so as to differently weighing  the contributions of a node's in-neighborhood and out-neighborhood.  
 For the sake of brevity here, we will refer to only one of the  formulations described in~\cite{TagarelliI13,Tagarelli14}, which is that based on the full \textit{in-out-neighbors-driven lurker ranking}, hereinafter named as   \textsf{LurkerRank} ($LR$). 

Given a directed social graph  $G  = \langle V, E \rangle$, where any edge $(u, v)$ means that $v$ is is ``consuming'' or ``receiving'' information from $u$, the  \textsf{LurkerRank}  $LR(v)$ score of  node  $v$ is defined as:
\begin{equation*}\label{eq:LR}
LR(v) = \alpha  [\mathcal{L}_{\mathrm{in}}(v)  \ (1+\mathcal{L}_{\mathrm{out}}(v))]  +   (1-\alpha)p(v)  
\end{equation*}
where $\mathcal{L}_{\mathrm{in}}(v)$ is the in-neighbors-driven lurking function:
\begin{equation*}
 \mathcal{L}_{\mathrm{in}}(v) =  \frac{1}{|\mathcal{N}^{out}_{v}|} \sum_{u \in \mathcal{N}^{in}_{v}} \frac{|\mathcal{N}^{out}_{u}|}{|\mathcal{N}^{in}_{u}|} LR(u) 
\end{equation*}  
and $\mathcal{L}_{\mathrm{out}}(v)$ is the out-neighbors-driven lurking function:
\begin{equation*}
\mathcal{L}_{\mathrm{out}}(v) =  \frac{|\mathcal{N}^{in}_{v}|}{\sum_{u \in \mathcal{N}^{out}_{v}} |\mathcal{N}^{in}_{u}|} \sum_{u \in \mathcal{N}^{out}_{v}} \frac{|\mathcal{N}^{in}_{u}|}{|\mathcal{N}^{out}_{u}|} LR(u)  
\end{equation*}  
where  $\alpha$ is a damping factor ranging within [0,1] (usually set to 0.85), and $p(v)$ is the value of the personalization vector, which is set to $1/|V|$ by default.   
To prevent zero or infinite ratios, the values of the in/out-neighborhood size   of a node  are Laplace add-one smoothed.  
As a result, the higher the $LR$ score of a node, the higher its likelihood to be regarded as a lurker in the network under study.

It should be noted that the actual meaning of  ``received information'' modeled by the links in the \textsf{LurkerRank} input graph can depend on the specific context of network analysis; in practice, it refers to either a social graph (i.e., a linked pair $(u,v)$  means that $v$ \textit{is follower} of $u$)  or an interaction graph (e.g., $v$ \textit{likes} or \textit{comments} $u$'s posts).   
  \textsf{LurkerRank} has been extensively evaluated on both scenarios~\cite{Tagarelli14,TagarelliI15}. 
 Nonetheless, although  both social and interaction relations are indicators of information consumption by users, the   information on interaction data that can be acquired from a real social network might be significantly sparse, and our context of study does not make an exception to this. Therefore, in this work    \textsf{LurkerRank} is applied to a followship graph, which corresponds to the Mastodon user networks defined in Section~\ref{sec:network-analysis} (with reversed edge-orientation).

 \end{document}